\documentclass[twocolumn]{aastex631}

\makeatletter
\let\frontmatter@title@above=\relax
\makeatother

\usepackage{booktabs} % nicer table rules
\usepackage{multirow} % multi-row table cells
\usepackage{soul} % highlighting (if needed)
\usepackage[normalem]{ulem} % underlining
\usepackage{bm}           % bold math symbols
\usepackage{placeins}     % \FloatBarrier
\graphicspath{ {./figs/} }

\begin{document}

\title{Open Science in Astrophysics: \\
 Citation Benefits of Open Code, Open Data, and Open Access}

\author{Parth Joshi}
% ORCID: 0009-0008-1404-1918
\affiliation{McWilliams Center for Cosmology and Astrophysics,
 Department of Physics, Carnegie Mellon University,
 Pittsburgh, PA 15213, USA}

\author{Rupert A.~C.~Croft}
\affiliation{McWilliams Center for Cosmology and Astrophysics,
 Department of Physics, Carnegie Mellon University,
 Pittsburgh, PA 15213, USA}

\correspondingauthor{Parth Joshi}
\email{ppjoshi@andrew.cmu.edu}

% Running head
\shorttitle{Open Science Citation Benefits in Astrophysics}
\shortauthors{Joshi \& Croft}

%\received{June 23, 2026}

% ============================================================
\begin{abstract}
We analyze the relationship between open-accessibility in data, code, and paper text in astrophysics using a sample of 53,194 peer reviewed papers published between January 2021 and April 2025, drawn from NASA’s Astrophysics Data System (ADS). 
We measure eleven quantities: open accessibility of text, open-code status, open-data status, number of grants received, code size, programming language, data repository size, citation count, number of authors, paper length, and publication date. 
We break down citation advantages based on six astrophysical sub-fields:  Solar System, Planet, Stellar, ISM, High Energy, and Galaxies\,+\,Cosmology, determined by keywords. 
This is accomplished by tuning a multivariate least-squares regression model alongside partial correlations and non-parametric tests to isolate the contribution of each facet of openness. After controlling for the aforementioned quantities, we find significant citation advantages associated with all three forms of openness: open data ($+32\%$, $p < 10^{-24}$), open access ($+26\%$, $p < 10^{-67}$), and open code ($+16\%$, $p = 0.003$).  The open-data citation advantage is present in all six sub-fields, and especially in Galaxies\,+\,Cosmology and ISM, which have the strongest cultures of sharing simulation outputs and observational data products. Open-code and open-data sharing rates are highest in Galaxies\,+\,Cosmology and HEA ($\sim$0.9\% and $\sim$2.9\%), reflecting their more developed community data infrastructure, and lowest in Solar System and ISM, where data is distributed on platforms not taken into account by this study.  Our findings support the long held notion that public access comes with concrete personal incentives for authors in terms of citations.

\end{abstract}

% Journal designation (for production use)
% \submitjournal{PASP}


% Fix for undefined currCollabLimit0 csname when no \collaboration is used
\makeatletter
\expandafter\xdef\csname currCollabLimit0\endcsname{9999}
\makeatother

% NOTE: \maketitle is NOT called explicitly here.
% AASTeX631 calls it automatically when the first \section{} is reached.

% ============================================================

\section{Introduction}
\label{sec:intro}
% ============================================================

Science is a field that would be impossible to further without relying on the contributions of others, whether that be through replicating their methods or using their data. However, traditionally research has been stuck behind paywalls, such as those incurred by journals or those required to store data \citep{Harnad2008, Laakso2011}. It is for that reason that open science practices, whereby authors can publish their results for all to read via open accessibility, is so impactful in accelerating research \citep{Tennant2016}.

While we know that open-accessibility is a social good due to how it strongly accelerates the overall pace of research in a field \citep{Piwowar2018}, it is more directly relevant to individual researchers and funding agencies to know the personal benefits for open access. If so, researchers have a stronger incentive to adopt open access practices - which would advance science as a whole. 

The largest personal benefit associated with open access would be in its scientific impact, which can be measured through bibliometrics. Citation count is a useful proxy for this measure \citep{Belter2015}. In theory, each citation acts as a vote of confidence from one paper to another \citep{Belter2015}. Metrics based on citation count are currently used to calculate journal impact factors and affect hiring decisions \citep{Hicks2015}.

%This makes it clear why 
Online availability has been historically associated with elevated citation rates across disciplines, and especially in physics \citep{Lawrence2001, Hajjem2006}. This effect was shown to not be due to selection bias, as \citet{Gargouri2010} found that it was present while controlling for mandation of open accessibility. Later reviews have confirmed a positive association, though the field has an effect on its magnitude \citep{Swan2010, Tennant2016, Piwowar2018}. 

Complementary to open accessibility is open data, whereby research datasets are made publicly available. This is done using repositories such as Zenodo \citep{zenodo} or Figshare \citep{figshare}. This lets other researchers further analyze the raw data that was collected and draw new conclusions from it. \citet{Piwowar2007} showed that in biology, sharing gene expression microarray data led to a 69\% increase in citations after controlling for journal impact factor, date of publication, and author country of origin. A similar trend, albeit generally weaker in magnitude, was found across  multiple disciplines in a large-scale analysis of papers published in PLoS \citep{Colavizza2020}. This is likely due to increased visibility, data reuse, and credibility signalling \citep{Piwowar2013}. The FAIR data principles, requiring research outputs to be Findable, Accessible, Interoperable, and Reusable have been widely adopted for open data sharing \citep{Wilkinson2016}.

Another facet of open science is open code, where the code used for a paper is also made publicly available. By publishing the code used to create data, researchers can build on work without having to reconstruct the data processing pipelines that were originally used. 
%Analysis of code sharing mandates %implemented at the journal \emph{Science}  %found that papers that shared code were far %easier to reproduce the results of when %compared to papers that hadn’t. 
In \citet{Maitner2024}, it was shown that code sharing practices did have a strong impact on citation rates in the field of ecology. This is in part likely explained and foreseen by earlier work by \citet{Stodden2018}, who analyzed code sharing mandates implemented at the journal \emph{Science} and found that a detailed analysis of papers whose authors shared code freely was far easier than for papers for whom the code was inaccessible. FAIR data principles and dedicated software citation standards \citep{Smith2016,Katz2021} have emerged to give code contributors formal credit. 

Astrophysics has been a relatively early and strong adopter of this practice of open-science - since 1991 it has been standard to upload research to \texttt{arXiv}, an open-access preprint server, before publishing it. In \citet{Piwowar2018}, it was shown that the text of 87\% of papers in the field were available as open access - the most of any scientific discipline analyzed. Furthermore, infrastructure such as NASA’s Astrophysics Data System (ADS) \citep{Kurtz2000, Henneken2012} and the Virtual Observatory \citep{Szalay2001, Hanisch2004} has made open-science the norm in the field. This is to the extent that NASA implemented the Scientific Information Policy for the Science Mission Directorate in 2022, which makes all NASA-funded publications and data open access \citep{NASA_SPD41}. There is a similar initiative in Europe that aims to accomplish this across multiple disciplines, called “Plan S” \citep{coalition2019}. 

Despite this extensive background, the exact current relationship between open science practices and citation impact in astrophysics has not yet been determined. Previous studies focused on a single facet of openness, using samples from before the current era of widespread data and code sharing, or leaving out important confounding variables such as the number of authors and paper length \citep{Piwowar2018,Dorch2015}. Without controlling for these variables, their effects can overshadow the effect of openness on citation count. In addition, sub-fields differ in reliance on open data and in citation cultures. For example, research in Galaxies and Cosmology involves large collaborations, and researchers produce and use shared datasets and simulation outputs for the sake of convenience \citep[e.g.\ IllustrisTNG, the Chandra Source Catalog;][]{Nelson2019,Evans2010}. However, Solar System research is done by on average by  smaller groups that use instrument-specific data, which may be proprietary. No previous study has taken into account differences between sub-fields in their models for the effect of open accessibility.

In this paper, we use modern statistical controls to analyze the effect of three types of open science: open access text, open data, and open code. We assembled a sample of 53,194 peer-reviewed astrophysics papers published between January 2021 and April 2025 to perform this analysis. We use multivariate regression, partial correlations, and non-parametric tests to control for eight distinct metadata variables, including astrophysical sub-field. Our aim is to determine quantitatively the effect of open-science practices on citation count in astrophysics.

% The paper is organized as follows.
% Section~\ref{sec:data} describes the dataset, the data sources used to
% assemble it, and the methods by which each variable was measured.
% Section~\ref{sec:analysis} presents our statistical framework, including
% the regression models and partial-correlation estimators.
% Section~\ref{sec:results} reports the main findings, broken down by
% openness type, sub-field, and auxiliary variables.
% Section~\ref{sec:discussion} discusses the interpretation of our results,
% potential sources of bias, and implications for the community.
% Section~\ref{sec:conclusions} summarises our conclusions.

% ============================================================
\section{Data}
\label{sec:data}
% ============================================================

To measure the effect of open-science practices on citation impact, we assembled a dataset of peer-reviewed astrophysics papers in June 2025 with eight associated metadata variables:
\begin{enumerate}
 \item Open-access text status, with paper length in characters;
 \item Open-code availability with code-repository size in bytes and
 programming language breakdown;
 \item Open-data availability, with data-repository size in bytes;
 \item Citation count;
 \item Number of authors;
 \item Number of grants acknowledged;
 \item Publication date; and
 \item Astrophysical subject area
\end{enumerate}

Rather than using the direct level of financial support a paper received, we used the number of grants it received as a proxy as there was no reliable machine-readable mapping of grant identifiers to awarded amounts for the range of funding agencies in our sample. However, this proxy is imperfect: a single large NASA
grant and five small travel awards are counted the same way. The implications of this are discussed in Section ~\ref{sec:discussion}.

The astrophysical subject area (item 8) plays a key 
role in our analysis, as different sub-fields have different norms around open science practices. For example, cosmology researchers who rely heavily on shared simulation outputs may be more likely to both share data and cite papers that engage in open science.

To assemble a complete dataset, we used several platforms, described in the following section.

% --- subsection: Data Sources ----------------------
\subsection{Data Sources}
\label{ssec:sources}

\subsubsection{NASA Astrophysics Data System}
\label{sssec:ads}

Paper metadata was primarily collected via NASA’s ADS \citep{Kurtz2000}, a portal that provides comprehensive bibliographic information for papers published in astronomy and astrophysics. The ADS API was queried for the fields listed in table ~\ref{tab:ads_params}. Of importance was the \texttt{property} field, the response to which encoded whether a paper was open access and whether it was peer-reviewed. We discarded all papers which were not peer-reviewed from our sample. The \texttt{keyword\_norm} field provided subject keywords for each paper we reviewed, which we used to assign each paper to one of six astrophysical sub-fields. The \texttt{links\_data} field was used to access all external URLs associated with each paper. This included links to the full text of the paper, code repositories, and data archives.

\begin{table}[t]
 \centering
 \setlength{\tabcolsep}{6pt}
 \renewcommand{\arraystretch}{1.4}
 \caption{Metadata fields requested from the NASA ADS API for each
 paper in the sample.}
 \label{tab:ads_params}
 \begin{tabular}{ll}
 \toprule
 Field & Description \\
 \midrule
 \texttt{bibcode} & ADS unique paper identifier \\
 \texttt{pubdate} & Publication date \\
 \texttt{citation\_count} & Number of citations received \\
 \texttt{citation} & List of citing bibcodes \\
 \texttt{author\_norm} & Authors in normalised format \\
 \texttt{links\_data} & External links (GitHub, Zenodo, etc.) \\
 \texttt{property} & Flags (open access, refereed, etc.) \\
 \texttt{doi} & Digital Object Identifier(s) \\
 \texttt{identifier} & Cross-identifiers (arXiv ID, etc.) \\
 \texttt{keyword\_norm} & ADS normalised subject keywords \\
 \bottomrule
 \end{tabular}
\end{table}

From ADS, we drew a random sample of 53,194 peer-reviewed papers published between January 1, 2021 and April 30, 2025. This four-year window was chosen specifically after the COVID-19 pandemic, as that had a significant impact on open science culture \citep{Casey2025effect}. A four year time period was necessary to provide a sufficient number of papers for statistical analysis. Additionally, we found that paper age was our single strongest predictor of citation count (Section \ref{sec:results}) - confining the sample to a narrower age range thus led to less variability.

\subsubsection{GitHub}
\label{sssec:github}

We identified code repositories primarily by searching the \texttt{links\_data} field of each ADS record for URLs pointing to \mbox{GitHub}\footnote{\url{https://github.com}}. Although GitHub is not the only platform researchers upload code to, we chose to focus on it because of its popularity, and assumed that papers linking to GitHub are a representative sample of papers with open code access. However, we do acknowledge that by restricting
to GitHub we missed a fraction of open code papers.

To retrieve data about a paper’s code, we queried the GitHub API to retrieve the repository size in bytes, as well as the proportion of the repository written in distinct programming languages. When a paper referenced multiple repositories, we combined the repository size and proportions as if they were one. The results were contained in the \texttt{ProgLangs} database.

\subsubsection{Zenodo}
\label{sssec:zenodo}

To find data sharing papers, we searched the \texttt{links\_data} field from the ADS query for URLs to Zenodo (\texttt{zenodo.org}).\footnote{\url{https://zenodo.org}} Zenodo is an open repository maintained by CERN, and is a widely used research data archive. We chose to use it because of its widespread use and popularity, like GitHub. Also like GitHub, we acknowledge that by focusing only on Zenodo for data sharing, we mislabel a fraction of open data papers. However, we believe that papers on Zenodo provide a representative sample of open data papers in astrophysics, as the platform is easily accessible and low-cost. 

\subsubsection{Crossref}
\label{sssec:crossref}

For data about grants, we used Crossref, a non-profit organization which maintains Digital Object Identifiers (DOIs) for literature. Its records include grant acknowledgement information for a large fraction of papers, with information about public funding agencies and identifiers for a large proportion of papers \citep{Lammey2020}. We queried the Crossref API for the DOIs of every paper in our sample, and recorded the number of distinct grants listed. We used grant count as a proxy for total funding, as Crossref does not record the contribution of each grant it records. Although the database provides broad coverage, we acknowledge that not all grants are disclosed in the metadata. Only grants which have been registered with Crossref are available. This leads to the data being  skewed, as larger public grants are much more likely to be registered with Crossref when compared to smaller private ones.

\subsubsection{Unpaywall}
\label{sssec:unpaywall}

To find the length of a paper in characters, we needed access to the full text of the paper. A link to the full-text PDF was sometimes provided by the ADS API, but not always - and the provided link sometimes did not resolve. To collect the full text of some papers that were not available via ADS, we made use of the Unpaywall API. Unpaywall\footnote{\url{https://unpaywall.org}} is an aggregate database of open-access articles with full-text available, created by Heather Piwowar and Jason Priem. If the full text of a paper was not found via Unpaywall or from ADS, it was excluded from analyses involving paper length. Closed access papers were also excluded from analyses involving paper length, as definitionally their text is not available. 

% --- subsection: Subject Classification --------------
\subsection{Subject Classification}
\label{ssec:subjects}

We classified each paper into one of six astrophysical sub-fields using the \texttt{keyword\_norm} parameter from ADS, which returns keywords available from standardized datasets - in particular, the Unified Astronomy Thesaurus (UAT), as defined by \citet{accomazzi2014unifiedastronomythesaurus}. However, we included the keywords from all datasets, as not every paper had UAT normalized keywords. We matched these against a priority ordered set of regular expressions (Table~\ref{tab:subjects}) corresponding to the categories commonly used in astronomical literature: Solar System, Planets, Stellar, interstellar medium (ISM), high energy astrophysics (HEA), and Galaxies \,+\, Cosmology. Papers whose keywords matched no patterns were placed in a separate “Other” category, while papers with no normalized keywords were placed in a different “No Keywords” category. 

\begin{table}[t]
 \centering
 \setlength{\tabcolsep}{5pt}
 \renewcommand{\arraystretch}{1.4}
 \caption{Subject area classification scheme based on ADS normalized
 keywords. Categories are listed in priority order, so that
 a paper matching multiple patterns is assigned to the
 highest-ranked category.}
 \label{tab:subjects}
 \begin{tabular}{lp{5.2cm}}
 \toprule
 Subject Area & Representative Keyword Patterns \\
 \midrule
 Solar System & \texttt{solar system}, \texttt{sun}, \texttt{interplanetary}, \texttt{comets}
 \texttt{solar wind}, \texttt{meteors}, \texttt{solar terrestrial} \\
 Planets & \texttt{planet}, \texttt{minor planets}, \texttt{asteroids},
 \texttt{exoplanet}, \texttt{protoplanet}\\
 Stellar & \texttt{stars}, \texttt{stellar}, \texttt{novae},
 \texttt{white dwarfs}, \texttt{neutron star}, \texttt{pulsars}, \texttt{supernovae} \\
 ISM & \texttt{ism}, \texttt{interstellar}, \texttt{nebul},
 \texttt{astrochemistry}, \texttt{masers},
 \texttt{molecular cloud}, \texttt{dust extinction} \\
 HEA & \texttt{x ray}, \texttt{gamma ray}, \texttt{black hole},
 \texttt{accretion}, \texttt{gravitational waves},
 \texttt{cosmic rays} \\
 Galaxies+Cosmo & \texttt{galax}, \texttt{cosmo}, \texttt{large scale
 structure}, \texttt{dark matter}, \texttt{quasar}, 
 \texttt{dark energy}
 \texttt{agn} \\
 \bottomrule
 \end{tabular}
\end{table}

Table~\ref{tab:sample_summary} gives the resulting sample sizes. A substantial fraction of papers (41\%) fell into the ``No Keywords'' category, reflecting the fact that normalized keyword assignment in ADS is not uniform across journals. A further 18\% were classified as ``Other'' (keywords present but not matching any of the six patterns defined above). These two categories together account for $\sim$59\% of the sample and are included in the overall regression analysis, but excluded
from sub-field comparisons.

% --- subsection: Sample Summary --------------------
\subsection{Sample Summary}
\label{ssec:summary}

\begin{table}[t]
  \centering
  \setlength{\tabcolsep}{4pt}
  \renewcommand{\arraystretch}{1.3}
  \caption{Paper sample by subject area and openness category.
           Code/Data = papers with a GitHub/Zenodo link.
           OA = open access.  Percentages are row fractions.}
  \label{tab:sample_summary}
  {\small
  \begin{tabular}{lrrrr}
    \toprule
    Subject & $N$ & OA & Code & Data \\
            &     & (\%) & (\%) & (\%) \\
    \midrule
    Solar System  &  4,302 & 78.5 & 0.5 & 1.8 \\
    Planets       &  1,506 & 77.9 & 0.7 & 2.0 \\
    Stellar       &  5,399 & 79.1 & 0.7 & 2.2 \\
    ISM           &  1,786 & 80.8 & 0.5 & 1.9 \\
    HEA           &  3,660 & 81.3 & 0.8 & 2.8 \\
    Gal.+Cosmo    &  5,155 & 82.1 & 0.9 & 2.9 \\
    No Keywords   & 21,870 & 80.3 & 0.6 & 2.0 \\
    Other         &  9,516 & 79.8 & 0.5 & 1.9 \\
    \midrule
    \textbf{Total} & \textbf{53,194} & \textbf{80.5} & \textbf{0.6} & \textbf{2.1} \\
    \bottomrule
  \end{tabular}}
\end{table}

Key properties of the full sample are summarized in
Table~\ref{tab:sample_stats}. The median citation count is 5 and the mean is 11.6, consistent with the highly skewed, approximately log-normal distribution typical of citation data \citep{Radicchi2008,Thelwall2016}. Open access accounts for 80.5\% of papers, reflecting the historically high open-access rate of the
astrophysics literature \citep{Piwowar2018}. Open code (0.6\%) and open data (2.1\%) are far rarer, which limits our statistical power, and even further when
we break results down by sub-field.

Looking across sub-fields in Table~\ref{tab:sample_summary}, some clear patterns stand out. The open-code and open-data rates are highest in Galaxies\,+\,Cosmology (0.9\% and 2.9\%) and HEA (0.8\% and 2.8\%), and lowest in Solar System (0.5\% and 1.8\%) and ISM (0.5\% and 1.9\%). This is consistent with the different research cultures in these sub-fields: cosmology and large-scale structure research routinely produces simulation data and survey catalogues that are shared through community repositories. On the other hand, Solar System research is often tied to a specific mission or instrument whose data are distributed through dedicated archives (e.g., NASA's Planetary Data System\footnote{\url{https://pds.nasa.gov}} ) rather than Zenodo. The open-access rates are similar across sub-fields, ranging from 78\% to 82\%, suggesting that
journal publishing model is not strongly tied to research topic.

These sub-field differences in code and data sharing rates mean that a single aggregate citation advantage figure could in principle be dominated by one or two sub-fields. We therefore present sub-field breakdowns
throughout the analysis and check whether the overall trends hold within each category.

\begin{table}[t]
 \centering
 \setlength{\tabcolsep}{8pt}
 \renewcommand{\arraystretch}{1.4}
 \caption{Summary statistics for the full paper sample ($N = 53{,}194$).}
 \label{tab:sample_stats}
 \begin{tabular}{lrr}
 \toprule
 Quantity & Median & Mean $\pm$ S.E. \\
 \midrule
 Citations & 5 & $11.6 \pm 0.09$ \\
 Number of authors & 4 & $6.4 \pm 0.05$ \\
 Paper length (chars)\tablenotemark{a}
 & 57{,}840 & $62{,}300 \pm 350$ \\
 Number of grants & 1 & $1.5 \pm 0.01$ \\
 Code size (bytes)\tablenotemark{b}
 & 4.5\,MB & $28.0 \pm 4.5$\,MB \\
 Data size (bytes)\tablenotemark{c}
 & 63\,MB & $840 \pm 85$\,MB \\
 \bottomrule
 \end{tabular}
 \tablenotetext{a}{Open-access papers only ($N = 42{,}833$).}
 \tablenotetext{b}{Papers with GitHub link only ($N = 337$).}
 \tablenotetext{c}{Papers with Zenodo link only ($N = 1{,}101$).}
\end{table}

\begin{figure*}[tp]
 \centering
 \includegraphics[width=\linewidth]{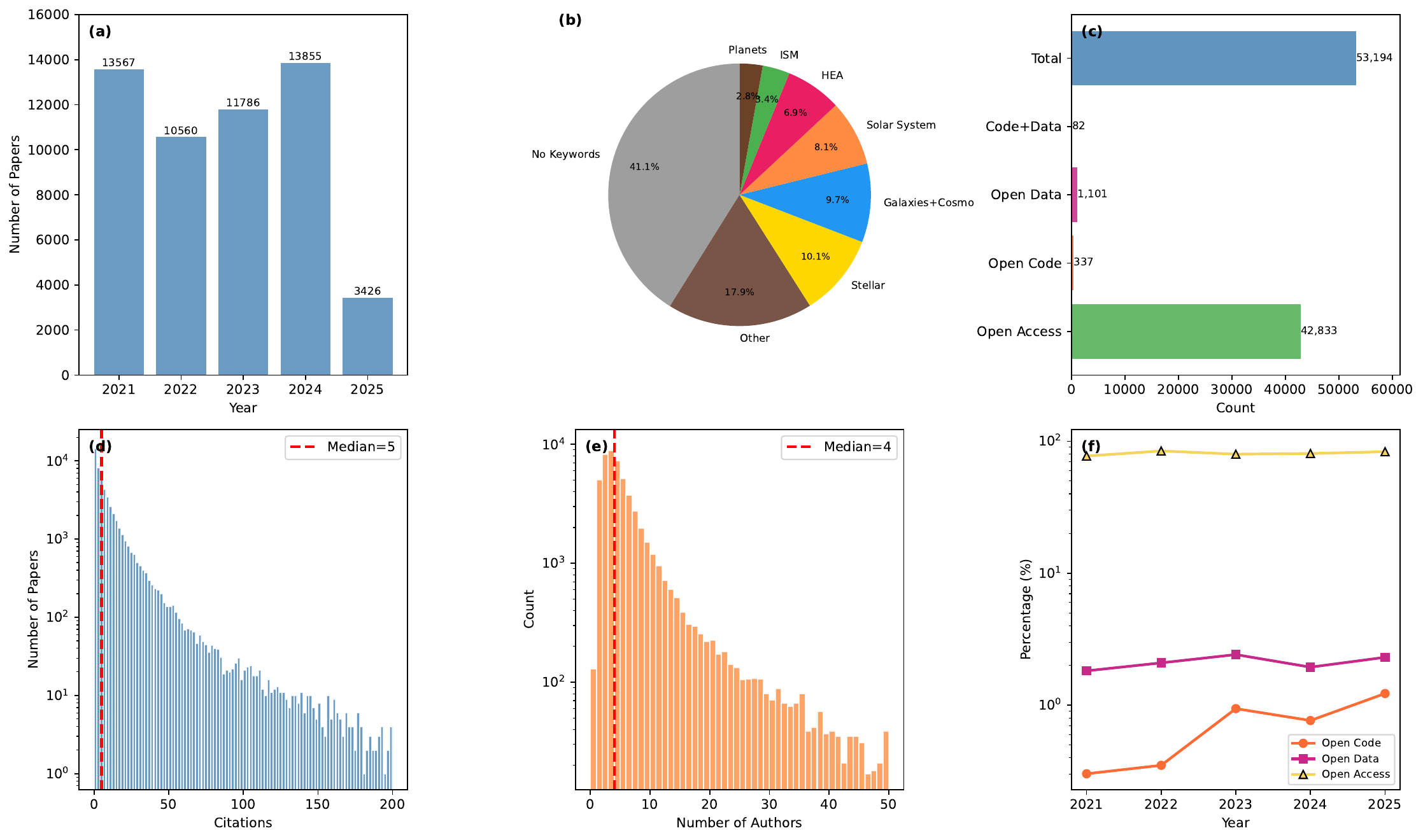}
 \caption{Dataset overview for the 53,194 astrophysics papers in our
 sample (2021--2025).
 \emph{(a)} Number of papers per publication year.
 \emph{(b)} Fraction of papers in each astrophysical sub-field
 (see Table~\ref{tab:subjects}).
 \emph{(c)} Counts in each openness category; ``Total'' is the
 full sample.
 \emph{(d)} Marginal citation distribution (clipped at 200
 for display); the dashed red line marks the median.
 Note the logarithmic vertical axis.
 \emph{(e)} Author-count distribution (clipped at 50),
 logarithmic vertical axis.
 \emph{(f)} Fraction of papers in each openness category per
 year, logarithmic vertical axis.}
 \label{fig:overview}
\end{figure*}

Figure~\ref{fig:overview} presents an overview of the dataset, showing the distribution of papers per year, the subject-area breakdown, the counts in each openness category, the marginal citation distribution,
the author-count distribution, and the time evolution of the open-access, open-code, and open-data fractions. The open-access fraction is relatively stable over the period, while the open-code and open-data fractions show a modest upward trend, consistent with the increasing
adoption of open-science practices encouraged by journal policies and funding-agencies \citep{coalition2019,NASA_SPD41,Wilkinson2016}.

% --- subsection: Variable Definitions ----------------
\subsection{Variable Definitions and Transformations}
\label{ssec:variables}

Some of the raw quantities we determined were highly skewed, such as citation count and author count. Throughout the analysis, unless stated otherwise, we take the log-transform of all variables - specifically $\log( 1 + x)$, which is defined for zero-valued observations and makes linear models more appropriate. We use the notation $\log N_{\rm auth}$ for $\log(1 + N_{\rm authors})$, $\log C$ for $\log(1 + N_{\rm citations})$, and so on.

We defined paper age  as the number of years elapsed between the publication date and the data-collection date (April~2025). For the regression analysis we use age in (fractional) years, ranging from $\sim$0.5 to $\sim$4.5 years across the sample. 
%We use age in years rather than months since a fixed epoch so that
%the regression coefficient has a directly interpretable scale.

The three openness variables \texttt{has\_code}, \texttt{has\_data}, and \texttt{open\_access} are binary indicators, as they are true or false. Their regression coefficients are therefore interpretable as the change in $\log C$ associated with the presence of that form
of openness, everything else being equal. The implied fractional change in citation count is $\exp(\hat\beta) - 1$, where $\hat\beta$ is the estimated coefficient.

% ============================================================
\section{Analysis}
\label{sec:analysis}
% ============================================================

Our analysis was done via two separate paths. In the first of these, we look at raw differences in citation count between papers with and without each form of openness using non-parametric tests so as to make no assumptions about the shape of the citation distribution (Section ~\ref{ssec:rawcomp}). In the second, to eliminate some confounding variables, we use multivariate regression and partial correlations to isolate the contribution of each openness variable while holding the others fixed (Section~\ref{ssec:regression}). We then apply these methods separately to papers from each astrophysical subfield (Section~\ref{ssec:subfield_analysis}). For open code papers, we also measure whether the programming language or the size of the repository makes a difference (Section~\ref{ssec:code_analysis}). Finally, we make special note of how the relationship between openness and citation rates has changed over time, and the role that grant funding in particular has as a confounding variable (Section~\ref{ssec:time_trends}).

Throughout we work with $\log C \equiv \log(1 + N_{\rm citations})$ as the
response variable rather than raw citation counts. As explained above, citation distributions
are highly skewed \citep{Thelwall2016} and the log transformation brings the residuals closer to
a normal distribution. This
improves the reliability of the standard errors and confidence
intervals. We report all regression results as both coefficients on the
log scale and as implied percentage changes in citation count,
$(\exp(\hat\beta)-1)\times 100$.

% --- 3.1 Raw comparisons ------------------------
\subsection{Raw Comparisons}
\label{ssec:rawcomp}

As a first step we compare citation distributions between papers with and without each form of openness, before applying any statistical controls. These raw comparisons are easy to read but can be misleading: if open-access papers happen to have more authors or to be older on average, a simple comparison will attribute to open access what is really driven by those other variables. We present them here to motivate the need for the multivariate approach that follows.

For each binary openness variable we split the sample into two groups and test whether their citation distributions differ using the two-sided Mann--Whitney $U$ test \citep{Mann1947}. This is a non-parametric rank-sum test that does not assume a normal distribution. We report the median citation count for each group, the ratio of the medians, and the $p$-value from the
$U$ test. Figure~\ref{fig:violin} shows violin plots of the full distributions on a log scale.

\begin{figure}[htb]
 \centering
 \includegraphics[width=\columnwidth]{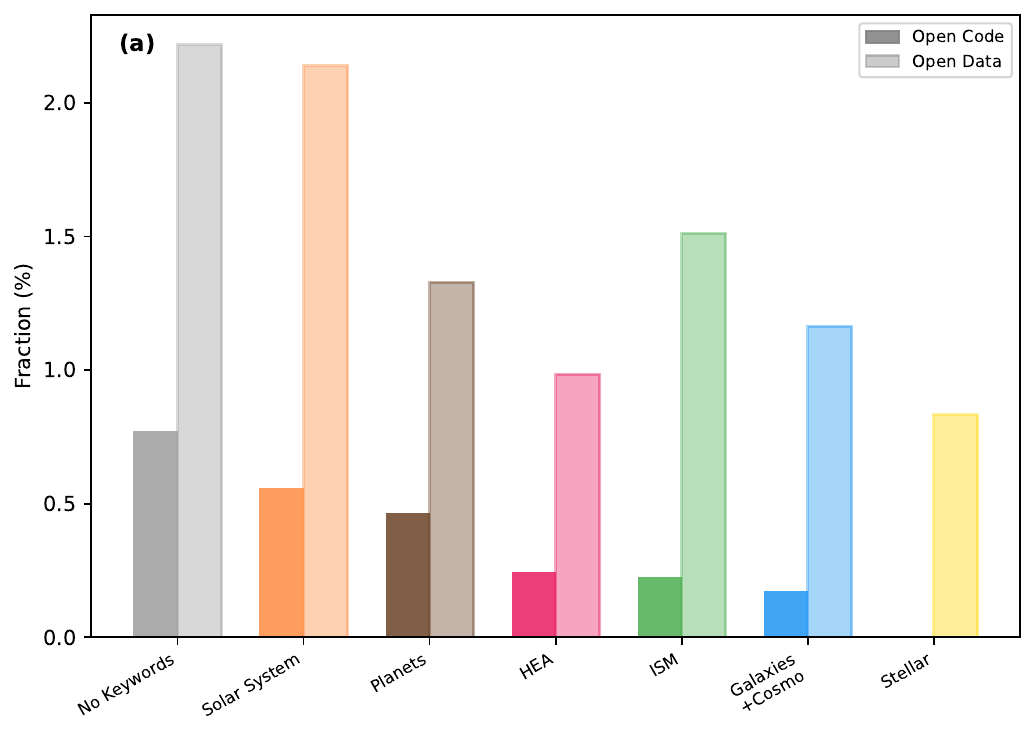}
 \caption{Open-code (filled bars) and open-data (outlined bars) sharing
 rates by astrophysical sub-field, sorted by code-sharing rate.
 See Section~\ref{ssec:disc_subfield} for discussion.}
 \label{fig:sharing_rates}
\end{figure}

Galaxies\,+\,Cosmology and HEA have the highest open-code and open-data sharing rates, while Solar System and ISM have the lowest (Figure~\ref{fig:sharing_rates}).

The raw comparisons seem to reveal a complication: papers with open code have \emph{lower} median raw citations (3 vs.\ 5) than papers without ($p < 10^{-4}$).
However, this is a direct result of age bias, as code-sharing papers in our sample tend
to be more recent (see panel (f) of Figure \ref{fig:overview}) and have simply had less time to accumulate citations. Open data and open access both show a positive raw difference, but the magnitude of the open-access effect is partly inflated by the fact that open-access papers in our sample are slightly older and have more authors on
average. The raw numbers are given in Table~\ref{tab:raw_comparison}. These results
make clear that controlling for paper age and author count is essential before
drawing any conclusions.

\begin{table}[t]
  \centering
  \setlength{\tabcolsep}{3pt}
  \renewcommand{\arraystretch}{1.3}
  \caption{Raw citation comparison by openness type.
           $N_+$/$N_-$: papers with/without that openness form;
           $\tilde{C}_+$/$\tilde{C}_-$: median citations;
           $p$: two-sided Mann--Whitney $U$ test.}
  \label{tab:raw_comparison}
  {\small
  \begin{tabular}{lrrrrl}
    \toprule
    Openness & $N_+$ & $N_-$ & $\tilde{C}_+$ & $\tilde{C}_-$ & $p$ \\
    \midrule
    Open Code   &    337 & 52,857 & 3 & 5 & $2\!\times\!10^{-5}$ \\
    Open Data   &  1,101 & 52,093 & 6 & 5 & $5\!\times\!10^{-3}$ \\
    Open Access & 42,833 & 10,361 & 6 & 1 & $<\!10^{-300}$ \\
    \bottomrule
  \end{tabular}}
\end{table}

% --- 3.2 Multivariate regression -------------------
\subsection{Multivariate Regression}
\label{ssec:regression}

To separate the effect of each openness variable from the other factors that
drive citation counts, we fit ordinary least-squares (OLS) regression models
with $\log C$ as the response. The full model is:

\begin{eqnarray}
 \log C_i & = & \alpha
   + \beta_1 \, \mathtt{code}_i
   + \beta_2 \, \mathtt{data}_i
   + \beta_3 \, \mathtt{oa}_i
   + \beta_4 \, \log N_{{\rm auth},i} \nonumber \\
   & & + \, \beta_5 \, \log L_i
   + \beta_6 \, \log G_i
   + \beta_7 \, A_i
   + \bm{\gamma}^\top \mathbf{s}_i
   + \varepsilon_i,
 \label{eq:model}
\end{eqnarray}

\noindent where $\mathtt{code}_i$, $\mathtt{data}_i$, and $\mathtt{oa}_i$
are binary indicators for open code, open data, and open access;
$N_{{\rm auth},i}$ is the number of authors; $L_i$ is the paper length in
characters (set to zero for closed-access papers where it is not openly
measurable); $G_i$ is the number of grants; $A_i$ is the paper age in years;
and $\mathbf{s}_i$ is a vector of sub-field indicator variables with coefficient
vector $\bm{\gamma}$. All log terms use $\log(1+x)$.

We fit two versions: a \emph{base model} without the sub-field indicators, and
a \emph{full model} that includes them. Comparing the models will let us determine how the relationship between openness and citation rate is affected by the subfield topic of a paper.

\begin{figure*}[tp]
 \centering
 \includegraphics[width=\linewidth]{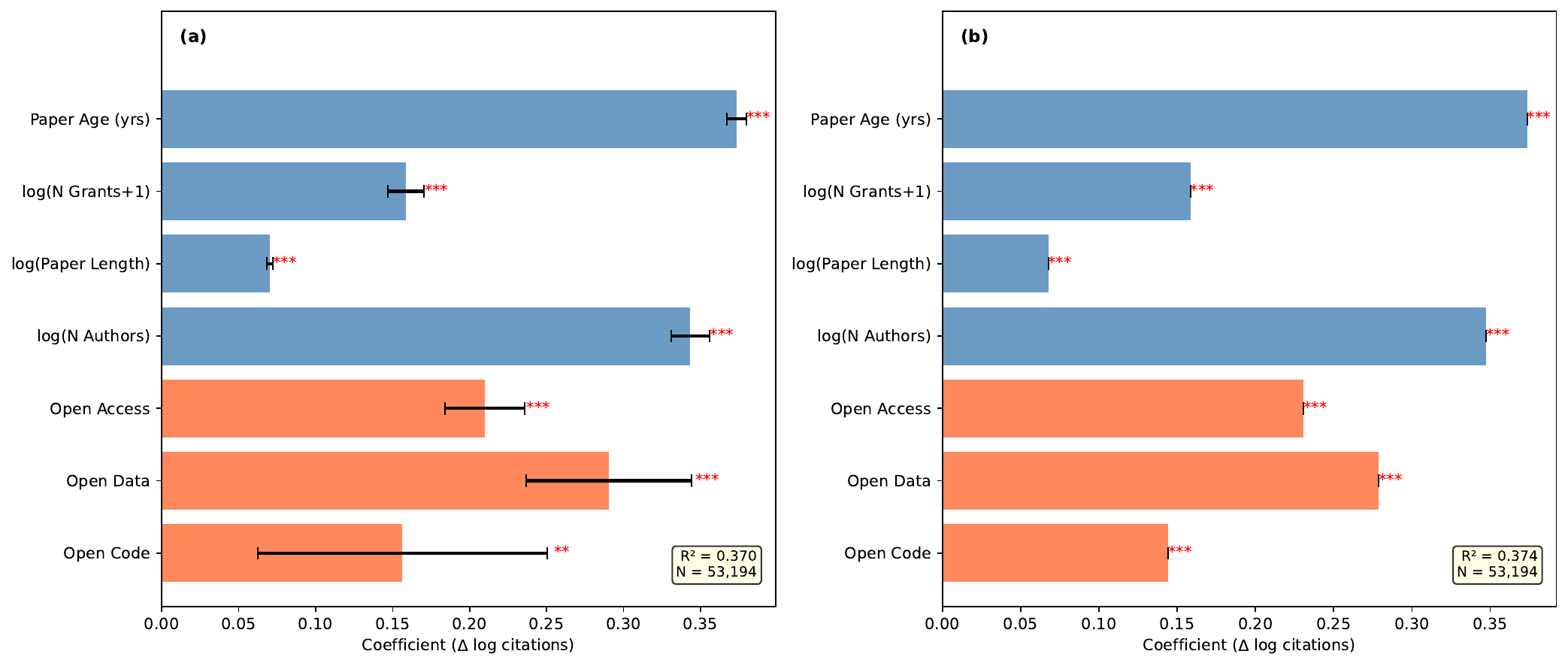}
 \caption{OLS regression coefficients on $\log(1+\mathrm{citations})$ with
 95\% confidence intervals (HC3 heteroskedasticity-robust standard
 errors). Orange bars are the three openness variables; blue bars
 are control variables.
 \emph{(a)} Base model without astrophysical sub-field fixed effects
 ($R^{2}=0.370$, $N=53{,}194$).
 \emph{(b)} Full model including sub-field indicator variables
 ($R^{2}=0.375$).
 Significance markers: $^{***}p<0.001$; $^{**}p<0.01$;
 $^{*}p<0.05$.}
 \label{fig:regression}
\end{figure*}

Because the residuals show mild heteroskedasticity as there is a larger scatter for highly cited papers, we use HC3 heteroskedasticity-robust standard
errors \citep{MacKinnon1985} throughout. These are more conservative than
classical standard errors and give more reliable $p$-values when the
equal-variance assumption is violated. Results are shown in
Figure~\ref{fig:regression} and Table~\ref{tab:regression}.

\begin{table}[t]
  \centering
  \setlength{\tabcolsep}{3pt}
  \renewcommand{\arraystretch}{1.3}
  \caption{
  Ordinary least-squares regression results for the citation model
           of Eq.~\ref{eq:model}, with $\log(1+\mathrm{citations})$ as the
           response. Each row gives the estimated coefficient $\hat\beta$, its
           HC3 heteroskedasticity-robust standard error (SE), and the two-sided
           $p$-value. Open Code, Open Data, and Open Access are binary
           indicators; the remaining rows are controls for author count, paper
           length, grant count, and age. The Base model excludes sub-field
           fixed effects, while the Full model includes a sub-field indicator
           for each astrophysical category. Positive coefficients correspond to
           higher citation counts.
}
  \label{tab:regression}
  {\small
  \begin{tabular}{lrrrr}
    \toprule
    & \multicolumn{2}{c}{Base} & \multicolumn{2}{c}{Full} \\
    \cmidrule(lr){2-3}\cmidrule(lr){4-5}
    Variable & $\hat\beta$ (SE) & $p$ & $\hat\beta$ (SE) & $p$ \\
    \midrule
    Open Code   & 0.156 (0.048) & 0.001 & 0.144 (0.048) & 0.003 \\
    Open Data   & 0.290 (0.027) & $<\!10^{-25}$ & 0.279 (0.027) & $<\!10^{-24}$ \\
    Open Access & 0.210 (0.013) & $<\!10^{-57}$ & 0.231 (0.013) & $<\!10^{-67}$ \\
    $\log N_{\rm auth}$ & 0.343 (0.006) & $<10^{-300}$ & 0.348 (0.006) & $<10^{-300}$ \\
    $\log L$ & 0.070 (0.001) & $<\!10^{-300}$ & 0.068 (0.001) & $<\!10^{-300}$ \\
    $\log G$ & 0.159 (0.006) & $<\!10^{-150}$ & 0.158 (0.006) & $<\!10^{-150}$ \\
    Age (yr) & 0.374 (0.003) & $<\!10^{-300}$ & 0.374 (0.003) & $<\!10^{-300}$ \\
    \midrule
    $R^2$ & \multicolumn{2}{c}{0.370} & \multicolumn{2}{c}{0.375} \\
    $N$   & \multicolumn{4}{c}{53,194} \\
    \bottomrule
  \end{tabular}}
\end{table}

The coefficients $\beta_1$, $\beta_2$, $\beta_3$ are our primary results.
Each gives the average change in $\log C$ when the corresponding openness
indicator is 1 rather than 0, holding all other predictors fixed. The
implied percentage changes in citation count are: open data $+32\%$, open
access $+26\%$, and open code $+16\%$. All three are statistically
significant at the 0.01 level in both model versions, showing that the
effects are not explained by sub-field differences.

% --- 3.2.1 Partial correlations -------------------
\subsubsection{Partial Correlations}
\label{sssec:partialcorr}

As a check on the regression results we also compute partial correlations
between each predictor and $\log C$. We removed the computed linear contribution of
the other controls, then computed the correlation. For a variable of
interest $x$ and response $y = \log C$, we find a fitting linear model, take the residuals $\tilde{x}$ and $\tilde{y}$, and report the Pearson correlation between them together with
its $p$-value. This gives a direct measure of the strength of each
association after the confounders have been accounted for
\citep{Kendall1948}.

\begin{figure}[htb]
 \centering
 \includegraphics[width=\columnwidth]{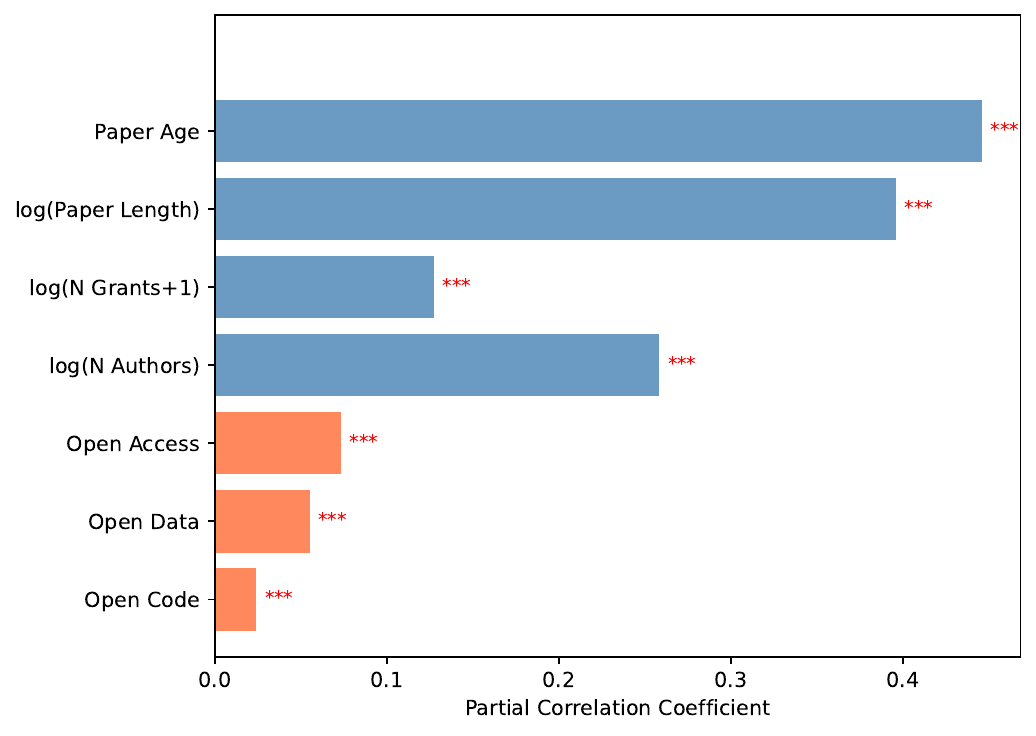}
 \caption{Partial correlations of each predictor with $\log(1 +
 \mathrm{citations})$ for the full dataset, after removing the
 linear contribution of the remaining control variables
 (number of authors, number of grants, paper length, paper age).
 Orange bars are the three openness variables; blue bars are
 control variables.
 Significance markers: $^{***}p < 0.001$; $^{**}p < 0.01$;
 $^{*}p < 0.05$.}
 \label{fig:partial_corr_full}
\end{figure}

Partial correlations for the full dataset are shown in Figure~\ref{fig:partial_corr_full}.
All three openness variables retain a significant positive partial correlation, confirming the regression result by an independent method.

% --- 3.2.2 Continuous predictors ------------------
\subsubsection{Continuous Predictors}
\label{sssec:continuous}

Alongside the binary openness variables, we measured the effect of continuous predictors on citation rate. Figure~\ref{fig:continuous} shows binned median citation counts as a function of each continuous predictor, with the 25th--75th percentile range shaded. We report Spearman rank correlations for these plots, since the relationships
are not necessarily linear on the log scale.

For code size and data size we restrict to the subsets of papers that actually share code or data, and ask whether, within those subsets, larger repositories are associated with more citations. We fit a separate
regression of the form:

\begin{equation}
 \label{eq:sizemodel}
 \log C_i = \alpha + \beta \, \log S_i
 + \beta_4 \log N_{{\rm auth},i}
 + \beta_6 \log G_i
 + \beta_7 A_i
 + \varepsilon_i,
\end{equation}

\noindent where $S_i$ is the repository size in bytes. This is fitted
separately for code papers ($N = 337$) and data papers ($N = 1{,}101$).

% --- 3.3 Sub-field breakdown ----------------------
\subsection{Sub-field Breakdown}
\label{ssec:subfield_analysis}

We are interested in checking whether the citation advantage of open science holds across all astrophysical sub-fields, or whether it is concentrated in one or two areas. We take two approaches.

First, we compute partial correlations of \texttt{code} and \texttt{data}
with $\log C$ within each sub-field (Figure~\ref{fig:partial_corr_subfield}).
This gives a visual summary of whether the overall trend is consistent across
sub-fields without requiring a separate regression for each.

Second, we show raw citation distributions for the four
openness-combination categories (no code/no data, code only, data only, code
and data) within each sub-field as box plots in
Figure~\ref{fig:subfield_breakdown}. Where sample sizes allow, we annotate
these with Mann--Whitney $p$-values comparing the
``no code/no data'' group against the code-sharing group.

Because open-code and open-data papers are rare (Table~\ref{tab:sample_summary}),
the per-sub-field samples are small and any regression coefficients have
large uncertainties. We therefore treat the sub-field results as
indicative rather than as precise measurements in their own right.

% --- 3.4 Code language analysis --------------------
\subsection{Code Language Analysis}
\label{ssec:code_analysis}

For the 337 papers for which we have access to the code we have the dominant language
of the repository. We ask two questions: (i) which languages are most common,
and does this vary by sub-field; and (ii) does the language relate to how many
citations the paper receives?

For question (i) we tabulate language counts by sub-field. For question (ii)
we compare median citation counts across languages and test for significant
differences using the Kruskal--Wallis test \citep{Kruskal1952}, a
non-parametric rank-based equivalent of the one-way $F$-test. We restrict to
languages with at least three papers to keep the per-cell counts manageable.%?
We caution that any apparent language-citation relationship could be driven
by sub-field confounding. For example if a particular language is used mainly in a
sub-field with higher baseline citations, that sub-field effect would show up
as a language effect. We therefore treat these results as exploratory.

Results are shown in Figure~\ref{fig:languages}.
Python is by far the most common language (panel~a).
The Kruskal--Wallis test finds no significant difference in citation count across languages ($p > 0.05$, panel~b);
the apparent variation is likely driven by sub-field confounding.

% --- 3.5 Time trends ---------------------------

\subsection{Time Trends and Grant Effects}
\label{ssec:time_trends}

We also check whether the citation advantage of open science has changed over
the 2021--2025 window, and whether grant funding interacts with openness.

For the time trend, we compute the difference in median citations between open
and closed papers in each calendar year (Figure~\ref{fig:advantage_time}).
The open-data advantage is positive in every year covered by the sample,
suggesting that the association between open data and higher citation counts is
not driven by a single year. The open-code advantage is also generally positive
but is noticeably noisier from year to year, which is consistent with the much
smaller number of code-sharing papers in each annual bin. Over this short
2021--2025 window we do not see strong evidence for a systematic increase or
decrease in the openness advantage, although the limited time baseline and small
sample sizes for code and data sharing mean that any conclusions about temporal
evolution should be treated cautiously.

We also examine the role of grant funding
(Figure~\ref{fig:grants_effects}). Papers acknowledging more grants tend to
have higher median citation counts, indicating that grant count is positively
associated with impact. In addition, papers with more grants are more likely to
share code and data, consistent with the idea that externally funded projects
may face stronger open-science expectations and may also have more resources to
prepare reusable public products. This makes grant count a potential confounder
when interpreting raw openness--citation relationships. We therefore check
whether grant count and openness have independent effects on citations, or
whether controlling for grants substantially reduces the apparent openness
advantage.

%\textcolor{red}{{\bf HERE!!!!!!!!!!}}
% ============================================================
\section{Results}
\label{sec:results}
% ============================================================

\subsection{Overall Citation Advantage of Open Practices}
\label{ssec:res_overall}

Table~\ref{tab:raw_comparison} showed that raw citation counts differ
between open and closed papers, but as discussed in
Section~\ref{ssec:rawcomp}, these differences are substantially
distorted by age and author-count biases.
\begin{figure*}[tp]
 \centering
 \includegraphics[width=0.98\linewidth]{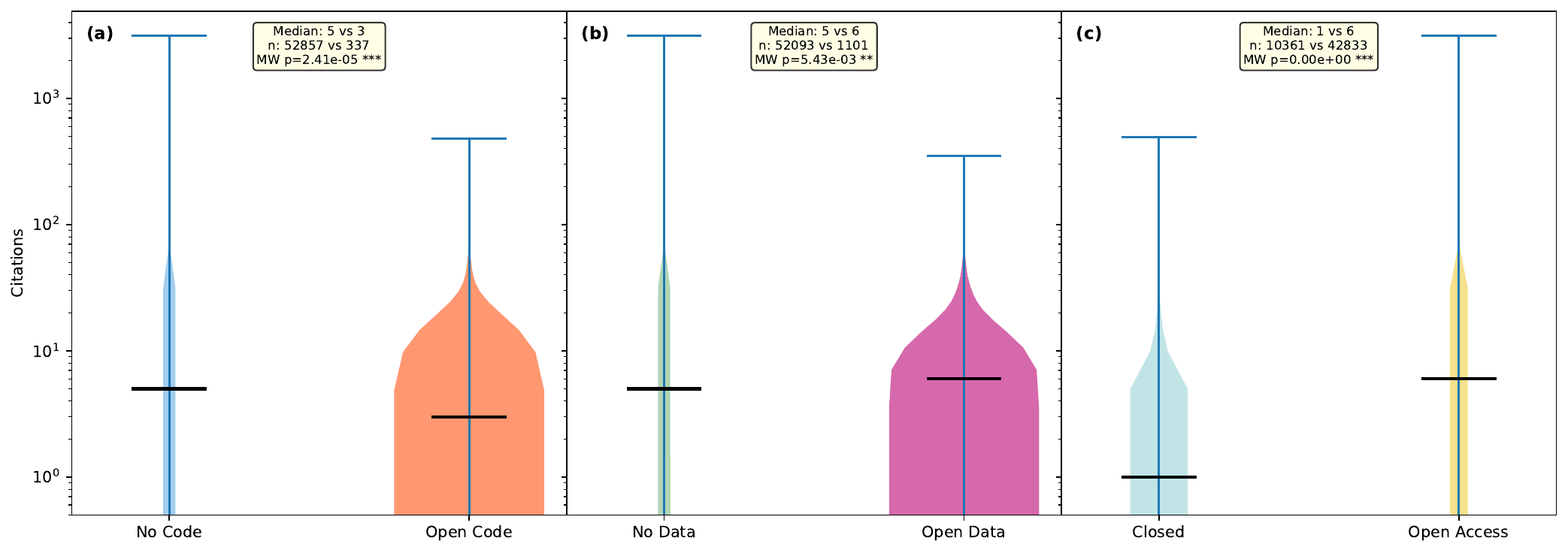}
 \caption{Raw citation distributions compared between open and closed papers.
 All panels use a logarithmic vertical axis.
 \emph{(a)} Violin plot comparing citation distributions for papers
 with and without open code.
 \emph{(b)} Same for open data.
 \emph{(c)} Same for open access vs.\ closed access.
 Horizontal bars mark group medians. Annotation in each panel gives
 sample sizes and the two-sided Mann--Whitney $U$ test $p$-value.}
 \label{fig:violin}
\end{figure*}

\begin{figure*}[tp]
 \centering
 \includegraphics[width=0.98\linewidth]{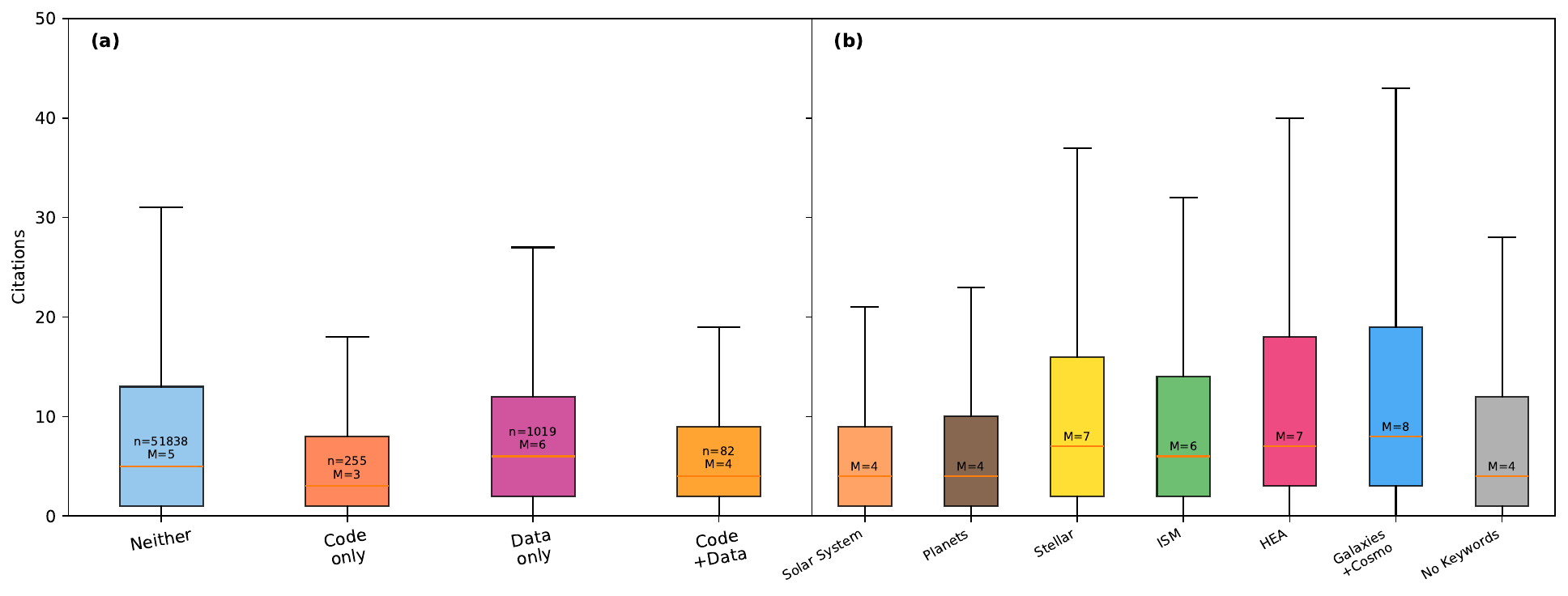}
 \caption{Citation distributions grouped by openness combination and
 by astrophysical sub-field.
 \emph{(a)} Box plots of citation counts for the four
 code/data combinations: papers with neither open
 code nor open data, code only, data only, and both.
 Boxes show the interquartile range; whiskers extend
 to 1.5 times the IQR. Medians ($M$) and sample
 sizes ($n$) are annotated.
 \emph{(b)} Same box-plot format showing citations across the
 seven subject categories (six science sub-fields plus
 the ``No Keywords'' group).
 Both panels share a common vertical axis with a maximum
 of 50 citations (99th percentile of the full sample).}
 \label{fig:combo_subject}
\end{figure*}

Table~\ref{tab:regression} gives the regression-corrected estimates.
After controlling for paper age, author count, paper length, and grant
support, all three forms of openness are associated with significantly
more citations.

Open data sharing is associated with a $32\%$ increase in citation
count ($\hat\beta = 0.290$, $p < 10^{-25}$). Open access is associated
with a $26\%$ increase ($\hat\beta = 0.210$, $p < 10^{-57}$). Open code
is associated with a $17\%$ increase ($\hat\beta = 0.156$, $p = 0.001$).
All three effects survive the addition of sub-field fixed effects in the
full model, where the estimates shift only slightly to $+32\%$, $+26\%$,
and $+16\%$ respectively (Table~\ref{tab:regression}). The stability of
the coefficients between the base and full models tells us that sub-field
differences in citation culture do not explain the openness advantage:
even within a given sub-field, open papers are cited more.

The largest single predictor in the model is paper age: on average, each additional
year since publication is associated with a $\sim 45\%$ increase in
citations ($\hat\beta = 0.374$). This is expected since citations accumulate
over time, and it is precisely why raw comparisons (which don't
account for the fact that code-sharing papers are disproportionately
recent) give a misleading picture. Author count is the second strongest
predictor ($\hat\beta = 0.343$ per unit of $\log N_{\rm auth}$, corresponding
to roughly $+40\%$ per doubling of author list), and the number of grants
is also significant ($\hat\beta = 0.159$, or $+17\%$ per doubling).

The model explains $R^2 = 0.370$ of the variance in $\log C$ without
sub-field terms, rising to $R^2 = 0.375$ with them. This is a modest improvement,
and consistent with sub-field differences being real but secondary to the
individual-paper predictors listed above.

Figure~\ref{fig:regression} shows the full set of coefficients with 95\%
confidence intervals. The three openness variables are all positive and
clearly separated from zero in both model versions.

\subsection{The Open-Code Reversal}
\label{ssec:res_reversal}

A striking feature of the raw data is that papers with open code have
a \emph{lower} median citation count than papers without (3 vs.\ 5,
Table~\ref{tab:raw_comparison}). This is counterintuitive given the
positive regression coefficient. It is however straightforward to 
understand this, being an effect of the fact that
code-sharing papers in our sample are systematically younger. In the
full dataset the mean age of a code-sharing paper is 1.8 years, compared
to 2.4 years for papers without code. Since age is the strongest
predictor of citation count, this implies younger code-sharing papers have simply had less time to accumulate citations.

Figure~\ref{fig:violin} (panel~a) illustrates
this clearly. On the log scale the two distributions overlap substantially,
and the difference in medians is driven by the tail of old, highly-cited
papers that mostly do not share code. Once age is
controlled for in the regression, the code-sharing advantage becomes
positive and significant.

%This reversal is a useful illustration of why controlling for confounders
%is essential in bibliometric studies. A paper that simply
%reported the raw medians would reach the wrong conclusion.

\subsection{Continuous Predictors}
\label{ssec:res_continuous}

\begin{figure*}[tp]
 \centering
 \includegraphics[width=\linewidth]{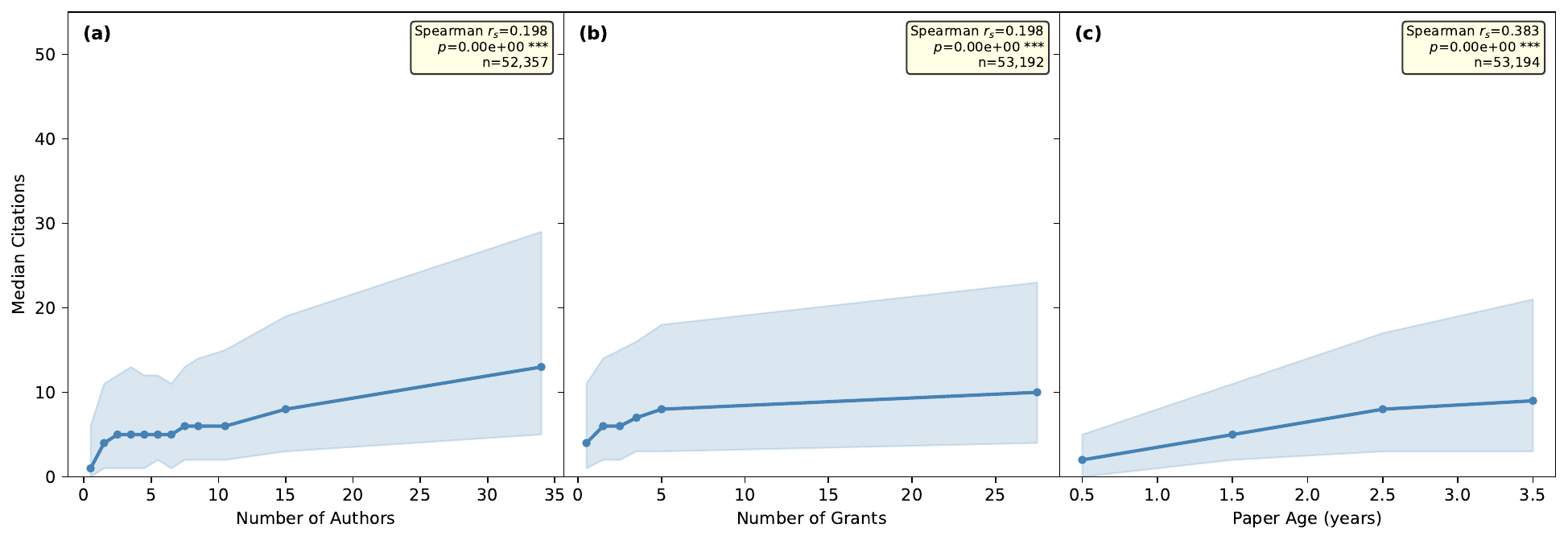}
 \caption{Binned median citations (solid line) as a function of each
 continuous predictor, with the 25th--75th percentile range
 shaded. The Spearman rank correlation $r_s$ and its $p$-value
 are given in each panel. For code size and data size, only
 papers that share code or data are included.}
 \label{fig:continuous}
\end{figure*}

Figure~\ref{fig:continuous} shows the relationship between each
continuous (non-binary) predictor and median citation count, after binning.

\subsubsection{Author Count}
The relationship between author count and citations is monotonically
positive up to at least $\sim 50$ authors, as can be seen in Figure~\ref{fig:continuous}. The Spearman correlation with raw citation count is among the highest of any predictor ($r_s \approx 0.35$, $p < 10^{-300}$). Large teams may produce higher-impact work for several reasons. First, they have more resources and wider expertise. They also draw on larger professional networks that generate citations. In practice, large collaborative papers often describe major datasets or software packages that accumulate many citations over time.

\subsubsection{Grant Count}
The number of grant acknowledgements shows a positive binned trend with
citations (Figure~\ref{fig:continuous}), with a Spearman correlation
of $r_s \approx 0.28$ ($p < 10^{-300}$). As noted in
Section~\ref{ssec:sources}, grant count is an imperfect proxy for
funding level, but it is the best available measure across the diversity
of funding agencies in our sample. The positive correlation is consistent
with the idea that better-resourced projects tend to produce more
influential work. It could also partly reflect the fact that
funded projects are more likely to appear in higher impact journals.

\subsubsection{Paper Length}
Longer papers are associated with more citations, with a Spearman
correlation of $r_s \approx 0.25$ ($p < 10^{-300}$) against raw
citation count. This is consistent with the intuition that more comprehensive papers
present more citable results. Because paper length can only be measured for open-access papers
(Section~\ref{ssec:disc_caveats}), this predictor is partially
collinear with open-access status in the regression. We set
$\log L = 0$ for closed-access papers, so the coefficient on
$\log L$ should be interpreted cautiously, as it likely absorbs some
of the open-access citation advantage.

\subsubsection{Repository Size}
Among the 337 papers that share code, the size of the GitHub repository
has no significant relationship to citation count after controlling for
age, author count, and grants ($\hat\beta = -0.021$, $p = 0.22$;
Equation~\ref{eq:sizemodel}). This suggests that the mere act of
sharing code is what matters for citations, not how much code is shared.

Among the 1,101 papers that share data on Zenodo, there is a small but
significant positive association between data repository size and citation
count ($\hat\beta = 0.018$ per log-byte, $p = 0.007$). A ten-fold
increase in data size corresponds to a $\sim 4\%$ citation increase.
The effect is small compared to the main data-sharing advantage ($+32\%$),
so it is a secondary consideration; but it suggests that larger 
 datasets attract modestly more citations, perhaps because
they are more comprehensive and broadly useful for reanalysis.

\subsection{Sub-field Differences}
\label{ssec:res_subfield}

\subsubsection{Baseline Citation Levels}
Sub-fields differ markedly in their typical citation counts even after
controlling for the individual-paper variables in
Equation~\ref{eq:model}. Relative to the Solar System baseline,
Galaxies\,+\,Cosmology papers receive $+10\%$ more citations
($\hat\gamma = 0.104$, $p < 10^{-7}$) and HEA papers $+8\%$ more
($\hat\gamma = 0.079$, $p < 10^{-6}$). By contrast, Planets papers
receive $-25\%$ fewer ($\hat\gamma = -0.291$, $p < 10^{-14}$), Stellar
$-7\%$ ($\hat\gamma = -0.075$, $p < 10^{-6}$), and ISM $-9\%$
($\hat\gamma = -0.093$, $p < 10^{-5}$).

These differences reflect the different citation cultures of each
sub-field. Galaxies\,+\,Cosmology and HEA involve large collaborations
and shared community products, such as simulation suites, survey databases,
and multi-mission source catalogues. These generate many citations
\citep{Nelson2019,Evans2010}. Planetary science, by contrast, tends
towards smaller teams and more instrument-specific results with a
narrower citing community. Figure~\ref{fig:combo_subject} (panel~b) shows that Galaxies\,+\,Cosmology has the highest median citation count across sub-fields, and Planets the lowest. The distributions are shown in that panel (bottom-centre
panel) illustrates the raw median citation distributions by sub-field.

\subsubsection{Open-Sharing Rates}
Open-code and open-data rates vary across sub-fields in a way that
mirrors the citation culture differences (Table~\ref{tab:sample_summary}).
Galaxies\,+\,Cosmology leads both lists, with 0.9\% of papers linking
to GitHub and 2.9\% linking to Zenodo. HEA is close behind (0.8\% code,
2.8\% data). Solar System (0.5\%, 1.8\%) and ISM (0.5\%, 1.9\%) have
the lowest sharing rates. This is consistent with the argument that
the Galaxies\,+\,Cosmology and HEA communities have more established
norms around making simulation and survey data publicly available, driven
partly by the scale of their data products and partly by a history of
community infrastructure such as the IllustrisTNG public release
\citep{Nelson2019} and the Chandra Source Catalog \citep{Evans2010}.

\begin{figure}[htb]
 \centering
 \includegraphics[width=\columnwidth]{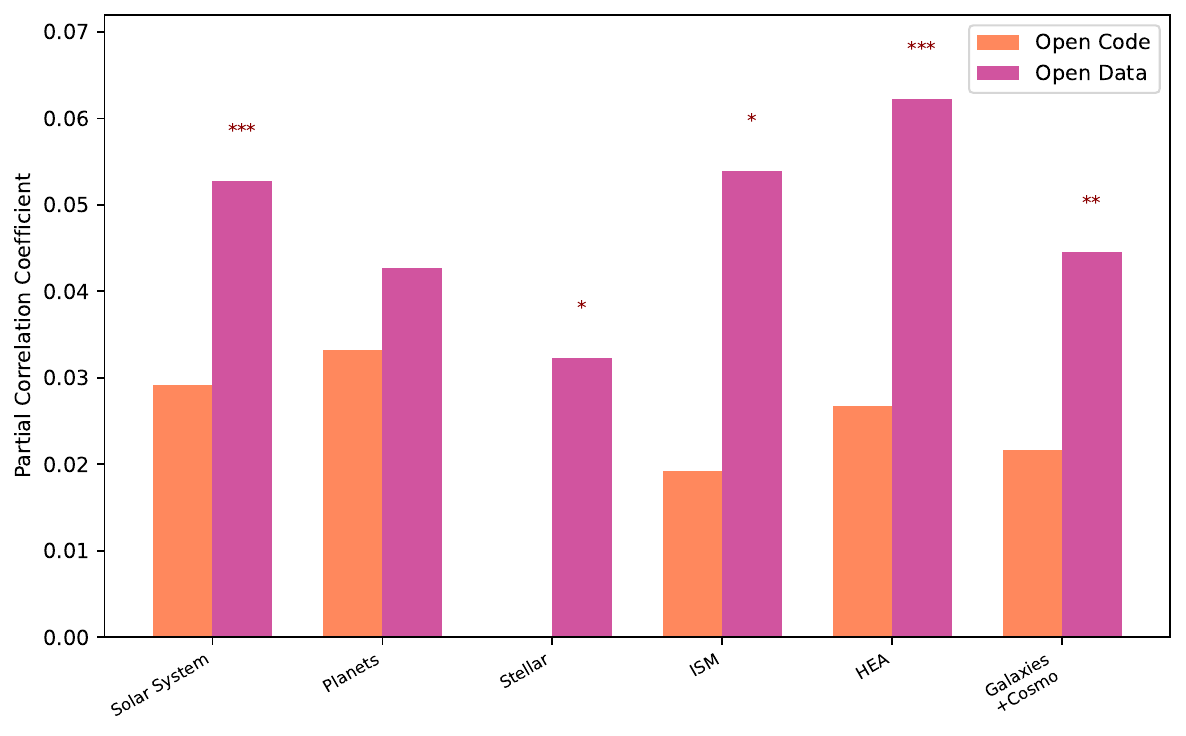}
 \caption{Partial correlations of open code (orange) and open data (pink)
 with $\log(1+\mathrm{citations})$ within each astrophysical
 sub-field, after controlling for author count, grant count,
 paper length, and paper age. Significance markers: $^{***}p < 0.001$; $^{**}p < 0.01$;
 $^{*}p < 0.05$.}
 \label{fig:partial_corr_subfield}
\end{figure}

\begin{figure*}[tp]
 \centering
 \includegraphics[width=\linewidth]{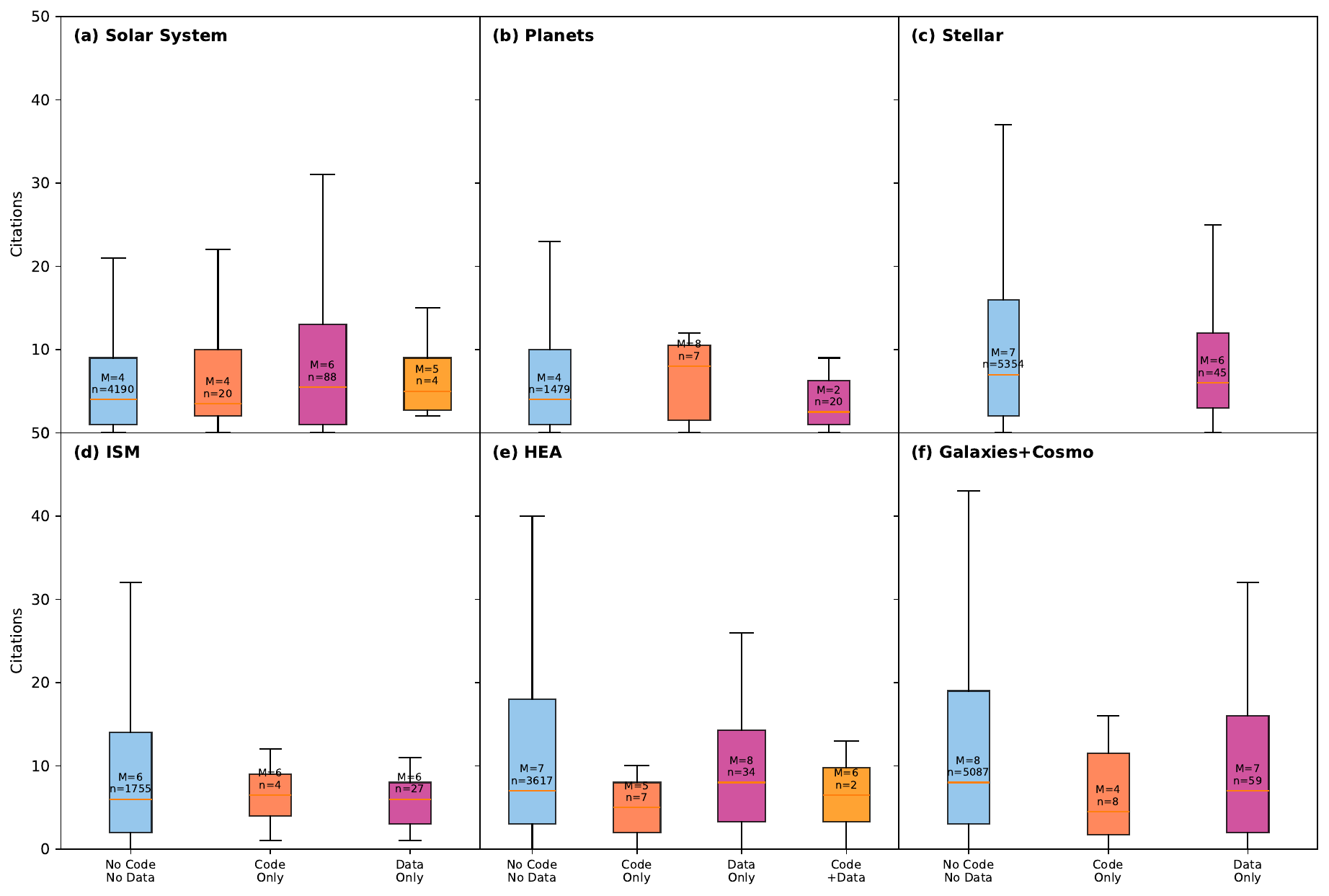}
 \caption{Citation distributions by openness combination within each
 astrophysical sub-field. Box shows the interquartile range;
 whiskers extend to 1.5 times the IQR; medians and sample sizes
 are annotated. The statistical annotation in each panel gives
 the Mann--Whitney $p$-value comparing papers with neither open
 code nor open data against those with open code.}
 \label{fig:subfield_breakdown}
\end{figure*}

\subsubsection{Open-Science Citation Advantage by Sub-field}
Figure~\ref{fig:subfield_breakdown} shows citation distributions split
by openness combination within each sub-field, and
Figure~\ref{fig:partial_corr_subfield} shows partial correlations of open code and open data within each sub-field. The open-data partial correlation is positive in all six sub-fields. The wide confidence intervals reflect the small number of code- and data-sharing papers per sub-field. The sub-field partial correlations
of open code and open data with $\log C$ within each sub-field.

The open-data partial correlation is positive in all six sub-fields,
though it is only statistically significant at $p < 0.05$ in the
full sample and in the larger sub-field groups (Galaxies\,+\,Cosmology,
Stellar, and HEA). The open-code partial correlation is similarly
positive in most sub-fields but has wide confidence intervals throughout,
reflecting the small number of code-sharing papers in each category.
No sub-field shows a clearly negative partial correlation for either
openness variable.  There is therefore no indication that
open sharing actively harms citation rates in any part of astrophysics.

The sub-fields where the open-data advantage appears largest in absolute
terms are ISM and Galaxies\,+\,Cosmology
%, both of which have strong
%cultures of sharing simulation outputs and observational data products.
In Solar System and Planets the data-sharing samples are too small for
reliable per-sub-field estimates, but the direction of the effect is
consistent with the overall positive trend.

\begin{figure*}[tp]
 \centering
 \includegraphics[width=\linewidth]{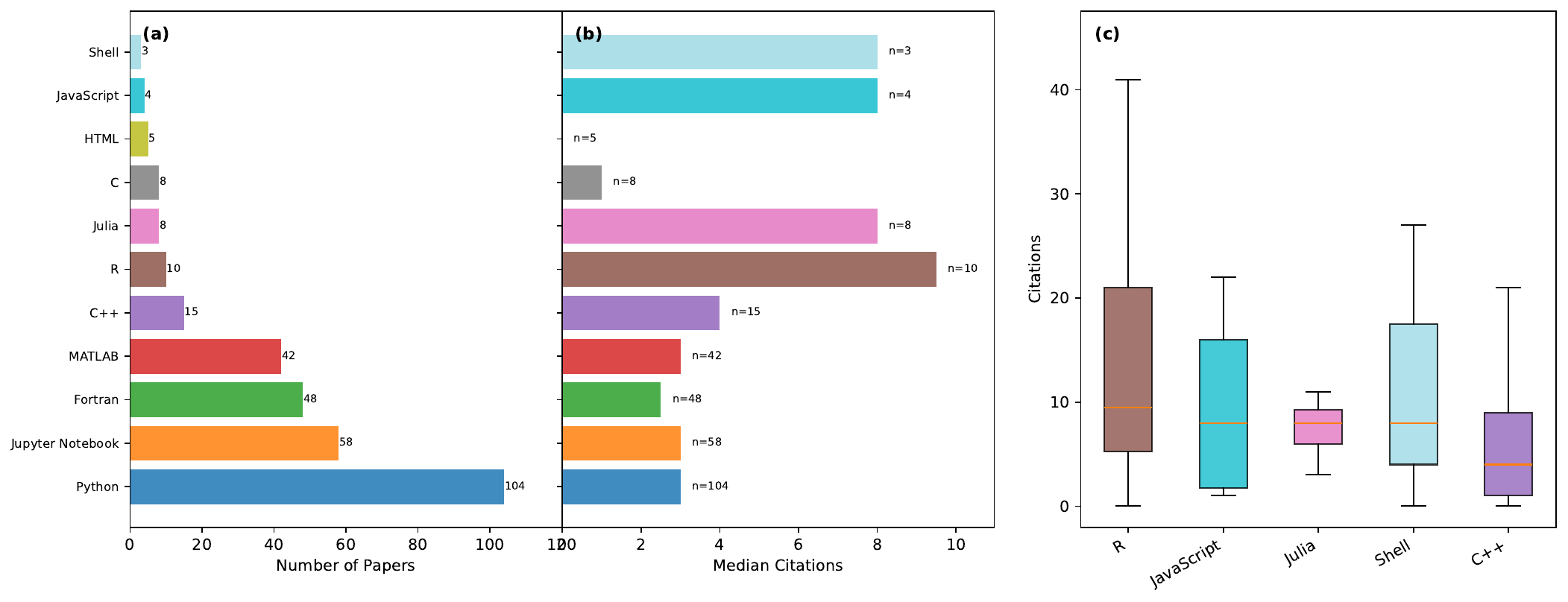}
 \caption{Programming language analysis for the 337 papers with a GitHub
 repository link. Colors are consistent across all three panels:
 each language has the same color in all panels.
 Languages with fewer than three papers are omitted.
 \emph{(a)} Number of papers per dominant language.
 \emph{(b)} Median citation count per language, sorted from
 highest to lowest.
 \emph{(c)} Box-plot citation distributions for the five most
 common languages.}
 \label{fig:languages}
\end{figure*}

\subsection{Code Language}
\label{ssec:res_language}

Figure~\ref{fig:languages} summarises the programming languages used
in code-sharing papers. Python is by far the most common language,
appearing in the majority of repositories, consistent with its dominant
role in modern astrophysics data analysis \citep{Momcheva2015}. Jupyter
Notebooks and R are the next most common, followed by Fortran and C/C++,
which tend to appear in papers involving large-scale numerical simulations.

Median citation counts vary across languages
(Figure~\ref{fig:languages}, panel~b), but the differences are
not statistically significant at the $p < 0.05$ level in the
Kruskal--Wallis test 
The apparent variation is likely driven by sub-field confounding: Fortran
and C/C++ papers disproportionately involve large simulation codes from
the Galaxies\,+\,Cosmology and HEA sub-fields, which have higher baseline
citation rates, while R is more common in statistical analysis papers
across a wider range of sub-fields.

\subsection{Time Trends}
\label{ssec:res_trends}

\begin{figure}[htb]
 \centering
 \includegraphics[width=\columnwidth]{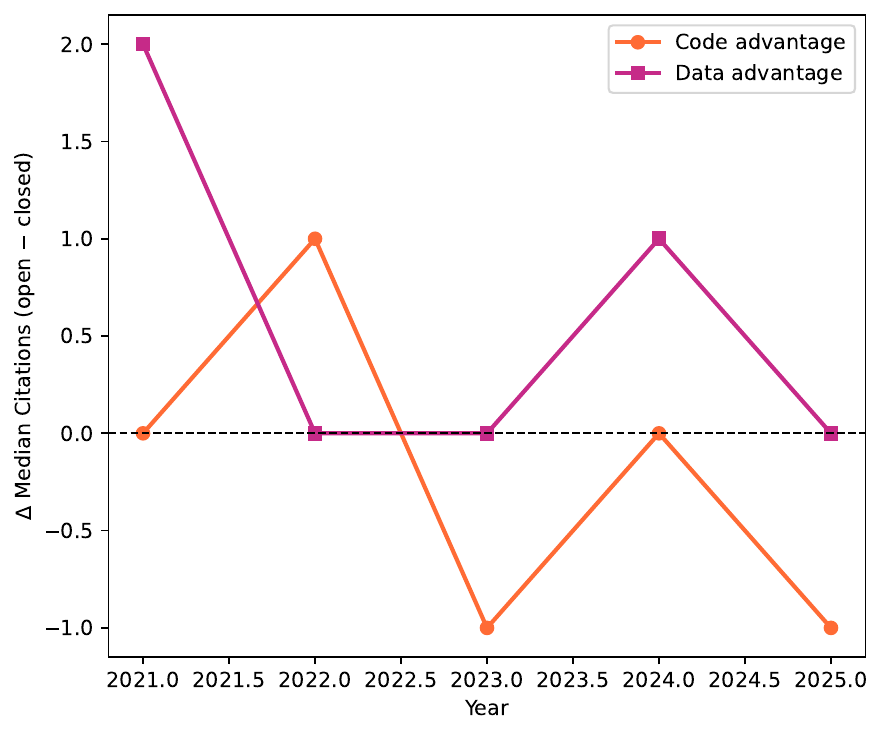}
 \caption{Difference in median citations between papers with and without open
 code or open data, shown as a function of publication year. The dashed line at
 zero is included for reference.}
 \label{fig:advantage_time}
\end{figure}

Figure~\ref{fig:advantage_time} shows the raw difference in median
citations between open and closed papers in each calendar year from
2021 to 2024 (2025 is excluded as the sample is incomplete). The
open-data advantage is positive in all years covered, ranging from
roughly $+1$ to $+3$ median citations. The open-code advantage is
noisier, reflecting the small number of code-sharing papers in any
single year ($\lesssim 100$ across the full 2021--2024 period), but
is also positive in most years. We caution that these are raw,
uncontrolled differences: year-to-year fluctuations are dominated
by the changing age distribution of papers within each year rather
than by genuine variation in the citation advantage. The short
time window prevents us from drawing firm conclusions about whether
the advantage is growing, stable, or shrinking over the period.

\subsection{Grant Funding and Openness}
\label{ssec:res_grants}

\begin{figure*}[tp]
 \centering
 \includegraphics[width=\linewidth]{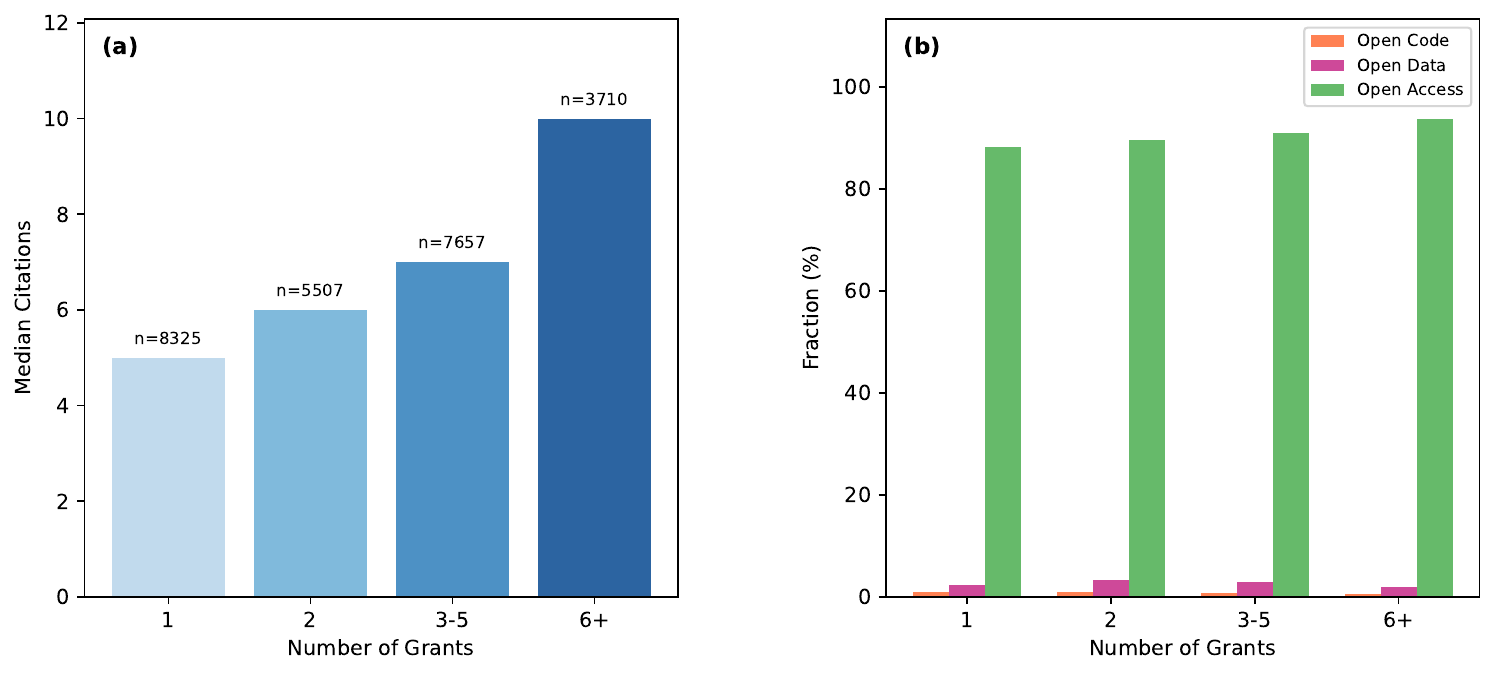}
 \caption{Grant-related trends in citation counts and openness.
 \emph{(a)} Median citation count as a function of the number of grants
 acknowledged, with sample sizes ($n$) annotated above the bars.
 \emph{(b)} Fraction of papers with open code, open data, or open access as a
 function of grant count.}
 \label{fig:grants_effects}
\end{figure*}

Figure~\ref{fig:grants_effects} (panel~b) shows that papers
acknowledging more grants are significantly more likely to share both
code and data.
Papers with five or more grants have open-code and open-data rates
roughly twice those of papers with no acknowledged grants. Open-access
rates show a similar but weaker trend. This is consistent with the
idea that funded projects have more resources to prepare data and code
for public release, and that funders increasingly require open practices
as a condition of awards \citep{NASA_SPD41,coalition2019}.

The citation advantage of open sharing persists after
controlling for grant count in the regression (Table~\ref{tab:regression}).
The openness coefficients change by less than 10\% when $\log G$ is
included versus excluded, indicating that the open-science citation
advantage is not simply a proxy for being well-funded.

% ============================================================
\section{Discussion}
\label{sec:discussion}
% ============================================================

\subsection{Interpreting the Citation Advantages}
\label{ssec:disc_interpret}

Our central finding is that open data, open access, and open code are
each independently associated with more citations, even after controlling
for paper age, author count, paper length, grant support, and
astrophysical sub-field. Before accepting these associations as causal,
it is worth asking what mechanisms could produce them and whether any
non-causal explanation remains plausible.

Three causal mechanisms are commonly proposed in the literature. First,
\emph{increased visibility}: papers that are freely available are
accessible to a wider audience, including researchers at institutions
without subscription access and those in lower-income countries
\citep{Tennant2016}. This mechanism is clearest for open
access, where the barrier removed is literal: a reader who cannot
download a paper cannot cite it. Second, \emph{reuse and reproducibility}:
papers that share data or code allow other researchers to re-analyse
the original dataset, extend the method, or validate the result, each of
which naturally generates a citation \citep{Piwowar2013,Stodden2018}.
This mechanism predicts that the data-sharing advantage should be larger
in sub-fields where community re-use of shared products is a normal part
of research practice, which is broadly consistent with our finding that
the advantage is most visible in Galaxies\,+\,Cosmology and HEA.
Third, \emph{quality signalling}: researchers who are confident in the
robustness of their work are more willing to share the underlying data
and code, so open sharing may act as a signal of quality that other
researchers respond to by citing more \citep{Piwowar2007}. This
mechanism is harder to test directly but is plausible given the positive
correlation between grant count (which is a proxy for external peer assessment
of quality) and sharing rates.

A fourth, non-causal explanation is \emph{residual confounding}: even
after controlling for the nine variables in our model, open papers may
differ systematically from closed papers in ways we have not measured.
The most likely candidate is journal prestige. High-impact journals
tend to require open access (either as policy or because authors of
high-profile papers can afford article processing charges), and papers
in high-prestige journals accumulate more citations independently of
their openness. We cannot fully rule out this explanation with the
available data, and we discuss it further in
Section~\ref{ssec:disc_caveats}.

\subsection{Comparison with Prior Work}
\label{ssec:disc_prior}

Our open-access citation advantage of $+26\%$ (controlled) is broadly
consistent with the range found in prior literature. \citet{Gargouri2010}
found advantages of $45$--$90\%$ in physics depending on the measure
used, though their sample predates widespread gold OA and their controls
are somewhat less comprehensive. \citet{Piwowar2018} found a $+18\%$
advantage in a large multi-disciplinary sample, somewhat lower than ours;
the difference may reflect the unusually high baseline OA rate in
astrophysics, which compresses the comparison group of closed papers
towards lower-prestige venues.

Our open-data advantage of $+32\%$ is in line with, though somewhat
lower than, the $+69\%$ reported by \citet{Piwowar2007} for microarray
data sharing in biomedicine. The gap is expected: the biomedical study
examined a highly specific and widely reused data type, while our sample
covers all astrophysics data on Zenodo, most of which will never be
re-analysed by a third party. \citet{Dorch2015} found a positive
data-sharing advantage in astronomy using a smaller and older sample,
consistent with our result. \citet{Colavizza2020} found a $+25\%$
advantage for data citation across disciplines in PLoS journals; our
higher figure may reflect the stronger data-sharing culture in
astrophysics relative to the average PLoS journal authors.

Ours appears to be the first study to report a controlled open-code
citation advantage in astrophysics. The $+16\%$ figure is modest but
statistically robust, and it persists across both model specifications.
This is consistent with the argument of \citet{Mislan2016} that code
sharing increases the reach and reuse of computational methods 
(their study was in ecology rather than astrophysics).

\subsection{Sub-field Differences}
\label{ssec:disc_subfield}

The sub-field analysis reveals differences in citation culture that are
worth discussing in their own right, independent of the openness question.
The $+10\%$ citation premium for Galaxies\,+\,Cosmology relative to
Solar System, and the $-25\%$ relative penalty for Planets, are large effects that
reflect real differences in how knowledge is built and cited across
astrophysical sub-disciplines.

Cosmology and large-scale structure involve a relatively small number of
landmark papers, such as simulation suites, survey catalogues, and parameter
estimation codes that accumulate very large citation counts because
they underpin a wide range of subsequent analyses
\citep{Nelson2019,Evans2010}. This ``blockbuster'' citation pattern
inflates the mean and median of the sub-field even for ordinary research
papers, because the community norm is to cite the relevant infrastructure
papers whenever results from those datasets are used. Planetary science,
by contrast, tends to involve instrument-specific datasets and results
that are not reused as widely outside the original team, producing a
more moderate citation distribution.

The pattern of sharing rates across sub-fields is similarly consistent
with community infrastructure. The IllustrisTNG collaboration
\citep{Nelson2019}, the EAGLE simulation \citep{Schaye2015}, and major
photometric and spectroscopic surveys all provide public data releases
that researchers deposit to Zenodo or link via ADS, driving up the
apparent Zenodo-sharing rate in Galaxies\,+\,Cosmology. Solar System
data, by contrast, are typically distributed through mission-specific
archives (e.g.\ the Planetary Data System) that do not appear in our
Zenodo-based measurement, so our open-data rate for that sub-field is
likely more strongly undercounted than for others.

\subsection{Limitations and Caveats}
\label{ssec:disc_caveats}

Several limitations of the present study deserve explicit discussion.

\subsubsection{Observational Design and Causality}
\label{sssec:disc_causal}

This is an observational study, so we cannot randomly assign papers to
open or closed conditions. This means that we can't establish causality from the
regression coefficients alone. The most plausible residual confounder,
as noted above, is journal prestige. A paper published in
\emph{Nature Astronomy} or \emph{Physical Review Letters} is both more
likely to be open access (because high-prestige work is more likely to
receive APC waivers or to be mandated open by funders) and more likely
to accumulate many citations (because the journal itself signals quality
and attracts readers). Without controlling for journal, our open-access
coefficient will absorb some of this effect. We note, however, that
the code and data sharing advantages are less likely to be explained by
journal prestige, since most journals do not yet have systematic
open-code or open-data policies that would create a prestige correlation
of this kind \citep{Stodden2018}.

\subsubsection{GitHub and Zenodo Coverage}
\label{sssec:disc_coverage}

We detect open code only via GitHub links in ADS records and open data
only via Zenodo links. Both platforms are popular in astrophysics but
neither is universal. Code is also shared via GitLab, Bitbucket,
personal websites, and institutional repositories, and data are deposited to
the Mikulski Archive for Space Telescopes (MAST), the NASA/IPAC Infrared
Science Archive (IRSA), the Planetary Data System (PDS), the ALMA
Science Archive, and many domain-specific databases. Our measured
open-code rate of $0.6\%$ and open-data rate of $2.1\%$ are therefore
lower bounds on the true fractions.

If papers that share code or data via other platforms are similar in
quality and citation profile to those that use GitHub and Zenodo, then
undercounting simply reduces our sample size for these groups without
biasing the estimated advantage. If, however, GitHub and Zenodo users
are systematically different (for example, if they are more likely to
be from well-resourced institutions or to work in certain sub-fields)
then our estimates could be biased. The sub-field differences in sharing
rates (Table~\ref{tab:sample_summary}) suggest some such heterogeneity
is present, and the Solar System case discussed above is a concrete
example where platform choice by the community is likely to cause
undercounting relative to other sub-fields.

\subsubsection{Grant Count as a Proxy}
\label{sssec:disc_grants}

We use the number of grant acknowledgements recorded by Crossref as a
proxy for the financial resources available to a research team. As noted
in Section~\ref{sec:data}, this proxy is imperfect in two ways. First,
Crossref's coverage is incomplete and varies by journal and funding
agency (some funders are systematically less likely to appear in
structured metadata). Second, the number of grants does not reflect their
size: a paper funded by a single large NASA grant may have more
resources than one with five small travel awards. These limitations mean
that our regression probably \emph{under}-controls for funding, so the
true openness coefficients after perfect funding control could be
somewhat smaller than those we report.

\subsubsection{Short Time Window and Citation Accumulation}
\label{sssec:disc_time}

Our sample covers only four and a half years (January 2021 to April
2025), and all citations were measured at a single point in time (April
2025). The youngest papers in the sample are therefore at most a few
months old at the time of citation measurement, which means their citation
counts reflect almost no accumulated impact. We mitigate this by
including paper age as a covariate and by log-transforming citations, but
the regression cannot fully capture the non-linear relationship between
age and citations for very young papers. Repeating the analysis with
a longer baseline, such as all papers from 2010 onwards would reduce
this problem substantially. This would however come at  the cost of 
relatively much smaller open-code and
open-data samples (since sharing practices were less common before 2015).

\subsubsection{Paper Length Collinearity}
\label{sssec:disc_length}

Paper length in characters is only available for open-access papers,
which means the $\log L$ term in the regression is partly collinear with
the $\mathtt{oa}$ indicator. Concretely, all closed-access papers have
$\log L = 0$ (the value we impute when length is unavailable), which
the regression cannot distinguish from papers that are genuinely very
short. This inflates the open-access coefficient slightly. The true
open-access advantage, net of any genuine paper-length effect, is
probably a few percentage points lower than our reported $+26\%$.

\subsection{Implications for Authors and Institutions}
\label{ssec:disc_implications}

Our results have practical implications at several levels.

For \emph{individual researchers}, the data suggest that the time
invested in preparing a public data release or a clean code repository
is likely repaid in increased citations. The $+32\%$ data-sharing advantage and
$+16\%$ code-sharing advantage are meaningful career effects given that
citation counts influence hiring, tenure, and grant decisions
\citep{Hicks2015,Waltman2016}. The fact that these advantages persist
after controlling for author count and grants means they are not simply
a by-product of working in large, well-funded collaborations.

For \emph{institutions and funding agencies}, the results support
existing mandates for open science \citep{NASA_SPD41,coalition2019}
and provide an additional, self-interested rationale for authors to
comply with them. The positive relationship between grant count and
sharing rates (Section~\ref{ssec:res_grants}) suggests that some of
this incentive is already operating: researchers with more funding are
more likely to share, consistent with funders' open-science requirements
having some effect.

For the \emph{astrophysics community specifically}, the sub-field
differences we document suggest that the benefits of open science are not
equally distributed. Galaxies\,+\,Cosmology and HEA researchers already
share data and code at higher rates and receive a citation premium that
partly reflects their existing data-sharing culture. Sub-fields with
lower sharing rates -- Solar System, ISM, Stellar -- may have more to
gain from developing community infrastructure analogous to the Virtual
Observatory \citep{Szalay2001} or the IllustrisTNG public release
\citep{Nelson2019}, both in terms of scientific productivity and
individual citation impact.

\section{Conclusions}
\label{sec:conclusions}
% ============================================================

We have assembled a sample of 53,194 peer-reviewed astrophysics papers
published between January 2021 and April 2025, drawn from the NASA
Astrophysics Data System, and measured nine bibliometric variables for
each paper: open-access status, open-code availability (via GitHub),
open-data availability (via Zenodo), citation count, number of authors,
paper length, number of grants (via Crossref), publication date, and
astrophysical sub-field. Using multivariate OLS regression with
heteroskedasticity-robust standard errors, partial correlations, and
non-parametric tests, we find the following.

\begin{enumerate}

\item {All three forms of openness are associated with more
citations, after controlling for other variables.}
Open data sharing is associated with a $+32\%$ citation increase
($\hat\beta = 0.290$, $p < 10^{-24}$), open access with $+26\%$
($\hat\beta = 0.210$, $p < 10^{-57}$), and open code with $+16\%$
($\hat\beta = 0.156$, $p = 0.001$). These estimates are stable to
the addition of astrophysical sub-field fixed effects, confirming that
the advantages are not simply an artefact of sub-field differences in
citation culture.

\item {Controlling for paper age is essential.}
Raw citation counts are misleading: code-sharing papers have
\emph{lower} raw median citations (3 vs.\ 5) because they are
systematically younger and have had less time to accumulate citations.
Once age is included in the model the code-sharing advantage flips
positive and significant. Paper age ($\hat\beta = 0.374$, $\sim 45\%$
per additional year) and author count ($\hat\beta = 0.343$, $\sim 40\%$
per doubling) are the two strongest predictors of citation count in our
dataset, which is why multivariate control is indispensable in any
comparison of open versus closed papers.

\item {Open-science sharing rates and citation levels both vary
across astrophysical sub-fields.}
Open-code and open-data rates are highest in Galaxies\,+\,Cosmology
(0.9\% and 2.9\%) and HEA (0.8\% and 2.8\%), reflecting the stronger
community infrastructure for sharing simulation outputs and survey
catalogues in those sub-fields. Solar System and ISM have the lowest
sharing rates, partly because mission data are often distributed through
dedicated archives that do not appear in our Zenodo-based measurement.
After controlling for individual-paper variables, Galaxies\,+\,Cosmology
papers receive $\sim 10\%$ more citations than the Solar System baseline,
while Planets papers receive $\sim 25\%$ fewer, reflecting real
differences in citation culture and community size.

\item {Repository size matters for data but not for code.}
Among papers that share code, the size of the GitHub repository has no
significant relationship to citation count ($p = 0.22$): what matters
is the act of sharing, not how much code is shared. Among papers that
share data on Zenodo, larger deposits are associated with a small but
significant citation premium ($+4\%$ per ten-fold increase in size,
$p = 0.007$), suggesting that broader datasets are modestly more useful
for reanalysis and thus attract more citations.

\item {Grant funding and openness are correlated but independent
predictors of citations.}
Papers with more grants are more likely to share code and data, and
both grant count and openness are independently associated with more
citations. The openness coefficients change by less than 10\% when
grant count is added to or removed from the model, indicating that the
open-science citation advantage is not simply a proxy for being
well-funded.

\end{enumerate}

These results should be interpreted with the caveats discussed in
Section~\ref{sec:discussion}. The study is observational, and residual
confounding by journal prestige(  which is the most plausible remaining
confounder) cannot be fully ruled out. The open-code and open-data
samples are small (337 and 1,101 papers respectively), limiting
statistical power for per-sub-field analysis. Our measurement of open
sharing is restricted to GitHub and Zenodo, so the true sharing fractions
are higher than we measure, and the degree of undercounting varies by
sub-field.

Taken together, however, the evidence is consistent: making papers, data, and code freely available is associated
with more citations in astrophysics, even after rigorous statistical
control. The magnitude of the effects, which vary from roughly $+16\%$ to $+32\%$
depending on the form of openness, is large enough to be practically
significant for individual researchers (in  hiring, tenure, and grant decisions \citep{Hicks2015}).
The astrophysics community already has much of the infrastructure needed
to make open sharing straightforward, from \texttt{arXiv} for preprints
to ADS for discovery to Zenodo and GitHub for data and code. Our results
suggest that using this infrastructure is not only good for science but
also good for the researchers who carry it out.

% ============================================================
\begin{acknowledgments}
This work was supported by a 2024-25 Seed Fund from the Block Center for
Technology and Society at Carnegie Mellon University.
We thank the NASA Astrophysics Data System team for making the ADS API
publicly available and for their continued maintenance of the
infrastructure on which this work depends. We are grateful to the
developers and maintainers of GitHub, Zenodo, Crossref, and Unpaywall
for providing open APIs that made it possible to assemble the dataset
described in Section~\ref{sec:data}. 
Parts of this manuscript were edited for clarity using Claude \citep{anthropic2025claude},
a large language model. All scientific content, interpretation,
and conclusions are solely the responsibility of the authors.
This research made use of the
Python open-source ecosystem, including 
NumPy \citep{numpy}, pandas \citep{pandas}, Matplotlib \citep{matplotlib},
SciPy \citep{scipy}, and statsmodels \citep{statsmodels}. 
The code used to generate the dataset is available at \url{https://github.com/hellothere98765/Code-For-Accessibility} \citep{mycode}. 
The datasets generated are available in Zenodo, at DOI \href{https://doi.org/10.5281/zenodo.20270191}{10.5281/zenodo.20270191} \citep{mydata}.
\end{acknowledgments}

\vspace{5mm}
\software{Python \citep{python3};
 NumPy \citep{numpy};
 pandas \citep{pandas};
 Matplotlib \citep{matplotlib};
 SciPy \citep{scipy};
 statsmodels \citep{statsmodels}}

\facility{ADS, \dataset[GitHub]{https://github.com},
 \dataset[Zenodo]{https://zenodo.org},
 Crossref, Unpaywall}

\bibliographystyle{aasjournal}
\bibliography{open_science_astro}

@misc{anthropic2025claude,
  author       = {Anthropic},
  title        = {Claude Sonnet 4.6 [Large language model]},
  year         = {2025},
  howpublished = {Anthropic PBC},
  note         = {Version \texttt{claude-sonnet-4-6},
                  \url{https://www.anthropic.com}}
}

@article{Tennant2016,
  author  = {Tennant, Jonathan P. and Waldner, Fran{\c{c}}ois and
             Jacques, Damien C. and Masuzzo, Paola and Collister, Lauren B.
             and Hartgerink, Chris H. J.},
  title   = {The academic, economic and societal impacts of Open Access:
             an evidence-based review},
  journal = {F1000Research},
  volume  = {5},
  pages   = {632},
  year    = {2016},
  doi     = {10.12688/f1000research.8460.3}
}

@article{Lawrence2001,
  author  = {Lawrence, Steve},
  title   = {Free online availability substantially increases a paper's impact},
  journal = {Nature},
  volume  = {411},
  pages   = {521},
  year    = {2001},
  doi     = {10.1038/35079151}
}

@article{Hajjem2006,
  author  = {Hajjem, Chawki and Harnad, Stevan and Gingras, Yves},
  title   = {Ten-Year Cross-Disciplinary Comparison of the Growth of Open
             Access and How It Increases Research Citation Impact},
  journal = {IEEE Data Engineering Bulletin},
  volume  = {28},
  number  = {4},
  pages   = {39--47},
  year    = {2006},
  note    = {arXiv:cs/0606122},
  url     = {http://sites.computer.org/debull/A05dec/hajjem.pdf}
}

@article{Gargouri2010,
  author  = {Gargouri, Yassine and Hajjem, Chawki and Lariviere, Vincent and
             Gingras, Yves and Carr, Les and Brody, Tim and Harnad, Stevan},
  title   = {Self-Selected or Mandated, Open Access Increases Citation Impact
             for Higher Quality Research},
  journal = {PLoS ONE},
  volume  = {5},
  number  = {10},
  pages   = {e13636},
  year    = {2010},
  doi     = {10.1371/journal.pone.0013636}
}

@article{Mislan2016,
 author = {Mislan, KAS and Heer, Jeffrey M and White, Ethan P},
 journal = {Trends in ecology \& evolution},
 number = {1},
 pages = {4--7},
 publisher = {Elsevier},
 title = {Elevating the status of code in ecology},
 volume = {31},
 year = {2016}
}

@article{Piwowar2018,
  author  = {Piwowar, Heather and Priem, Jason and Larivi{\`e}re, Vincent and
             Alperin, Juan Pablo and Matthias, Lisa and Norlander, Bree and
             Farley, Ashley and West, Jevin and Haustein, Stefanie},
  title   = {The state of {OA}: a large-scale analysis of the prevalence and
             impact of Open Access articles},
  journal = {PeerJ},
  volume  = {6},
  pages   = {e4375},
  year    = {2018},
  doi     = {10.7717/peerj.4375}
}

@misc{Swan2010,
  author       = {Swan, Alma},
  title        = {The {Open Access} Citation Advantage: Studies and Results to Date},
  year         = {2010},
  howpublished = {Technical Report, School of Electronics \& Computer Science,
                  University of Southampton},
  url          = {https://eprints.soton.ac.uk/268516/}
}

@article{Harnad2008,
  author  = {Harnad, Stevan and Brody, Tim and Vall{\`e}res, Fran{\c{c}}ois and
             Carr, Les and Hitchcock, Steve and Gingras, Yves and Oppenheim,
             Charles and Hajjem, Chawki and Hilf, Eberhard R.},
  title   = {The Access/Impact Problem and the Green and Gold Roads to Open
             Access: An Update},
  journal = {Serials Review},
  volume  = {34},
  number  = {1},
  pages   = {36--40},
  year    = {2008},
  doi     = {10.1016/j.serrev.2007.12.005}
}

@article{Laakso2011,
  author  = {Laakso, Mikael and Welling, Patrik and Bukvova, Helena and
             Nyman, Linus and Bj{\"o}rk, Bo-Christer and Hedlund, Turid},
  title   = {The Development of Open Access Journal Publishing from 1993 to 2009},
  journal = {PLoS ONE},
  volume  = {6},
  number  = {6},
  pages   = {e20961},
  year    = {2011},
  doi     = {10.1371/journal.pone.0020961}
}

@article{figshare,
author = {Thelwall, Mike and Kousha, Kayvan},
year = {2015},
month = {12},
pages = {},
title = {Figshare: A universal repository for academic resource sharing?},
volume = {40},
journal = {Online Information Review},
doi = {10.1108/OIR-06-2015-0190}
}

@article{Piwowar2007,
  author  = {Piwowar, Heather A. and Day, Roger S. and Fridsma, Douglas B.},
  title   = {Sharing Detailed Research Data Is Associated with Increased
             Citation Rate},
  journal = {PLoS ONE},
  volume  = {2},
  number  = {3},
  pages   = {e308},
  year    = {2007},
  doi     = {10.1371/journal.pone.0000308}
}

@article{Piwowar2013,
  author  = {Piwowar, Heather and Vision, Todd J.},
  title   = {Data reuse and the open data citation advantage},
  journal = {PeerJ},
  volume  = {1},
  pages   = {e175},
  year    = {2013},
  doi     = {10.7717/peerj.175}
}

@article{Dorch2015,
  author  = {Dorch, Bertil F. and Drachen, Thore M. and Ellegaard, Ole},
  title   = {The data sharing advantage in astrophysics},
  journal = {Proceedings of the International Astronomical Union},
  volume  = {11},
  number  = {A29A},
  pages   = {172--175},
  year    = {2015},
  doi     = {10.1017/S1743921316002696}
}

@article{Colavizza2020,
  author  = {Colavizza, Giovanni and Hrynaszkiewicz, Iain and Staden, Isla
             and Whitaker, Kirstie and McGillivray, Barbara},
  title   = {The citation advantage of linking publications to research data},
  journal = {PLoS ONE},
  volume  = {15},
  number  = {4},
  pages   = {e0230416},
  year    = {2020},
  doi     = {10.1371/journal.pone.0230416}
}

@article{Maitner2024,
author = {Maitner, Brian and Santos Andrade, Paul Efren and Lei, Luna and Kass, Jamie and Owens, Hannah L. and Barbosa, George C. G. and Boyle, Brad and Castorena, Matiss and Enquist, Brian J. and Feng, Xiao and Park, Daniel S. and Paz, Andrea and Pinilla-Buitrago, Gonzalo and Merow, Cory and Wilson, Adam},
title = {Code sharing in ecology and evolution increases citation rates but remains uncommon},
journal = {Ecology and Evolution},
volume = {14},
number = {8},
pages = {e70030},
doi = {https://doi.org/10.1002/ece3.70030},
url = {https://onlinelibrary.wiley.com/doi/abs/10.1002/ece3.70030},
eprint = {https://onlinelibrary.wiley.com/doi/pdf/10.1002/ece3.70030},
note = {e70030 ECE-2023-11-02004.R2},
year = {2024}
}

@article{Stodden2018,
  author  = {Stodden, Victoria and Seiler, Jennifer and Ma, Zhaokun},
  title   = {An empirical analysis of journal policy effectiveness for
             computational reproducibility},
  journal = {Proceedings of the National Academy of Sciences},
  volume  = {115},
  number  = {11},
  pages   = {2584--2589},
  year    = {2018},
  doi     = {10.1073/pnas.1708290115}
}

@article{Smith2016,
  author  = {Smith, Arfon M. and Katz, Daniel S. and Niemeyer, Kyle E. and
             {FORCE11 Software Citation Working Group}},
  title   = {Software citation principles},
  journal = {PeerJ Computer Science},
  volume  = {2},
  pages   = {e86},
  year    = {2016},
  doi     = {10.7717/peerj-cs.86}
}

@article{Katz2021,
  author  = {Katz, Daniel S. and Gruenpeter, Marc and Honeyman, Tom},
  title   = {Taking a fresh look at {FAIR} for research software},
  journal = {Patterns},
  volume  = {2},
  number  = {3},
  pages   = {100222},
  year    = {2021},
  doi     = {10.1016/j.patter.2021.100222}
}

@article{Wilkinson2016,
  author  = {Wilkinson, Mark D. and Dumontier, Michel and Aalbersberg,
             IJsbrand Jan and Appleton, Gabrielle and Axton, Myles and
             Baak, Arie and Blomberg, Niklas and Boiten, Jan-Willem and
             da Silva Santos, Luiz Bonino and Bourne, Philip E. and
             others},
  title   = {The {FAIR} Guiding Principles for scientific data management
             and stewardship},
  journal = {Scientific Data},
  volume  = {3},
  pages   = {160018},
  year    = {2016},
  doi     = {10.1038/sdata.2016.18}
}

@misc{NASA_SPD41,
  author       = {{NASA}},
  title        = {Scientific Information Policy for the Science Mission
                  Directorate ({SPD-41a})},
  year         = {2022},
  howpublished = {\url{https://science.nasa.gov/researchers/science-data/science-information-policy}},
  note         = {Accessed: 2025}
}

@misc{coalition2019,
  author       = {{cOAlition S}},
  title        = {Accelerating the transition to full and immediate {Open Access}
                 to scientific publications ({Plan S})},
  year         = {2019},
  howpublished = {Science Europe},
  url          = {https://www.coalition-s.org/plan-s-principles/}
}

@article{Kurtz2000,
  author  = {Kurtz, Michael J. and Eichhorn, Guenther and Accomazzi, Alberto
             and Grant, Carolyn S. and Murray, Stephen S. and Watson, Joyce M.},
  title   = {The {NASA} Astrophysics Data System: Overview},
  journal = {Astronomy \& Astrophysics Supplement Series},
  volume  = {143},
  pages   = {41--59},
  year    = {2000},
  doi     = {10.1051/aas:2000170}
}

@inproceedings{Henneken2012,
  author    = {Henneken, Edwin A. and Accomazzi, Alberto},
  title     = {Linking to Data: Effect on Citation Rates in Astronomy},
  booktitle = {Astronomical Data Analysis Software and Systems {XXI}},
  series    = {ASP Conference Series},
  volume    = {461},
  pages     = {763--766},
  year      = {2012},
  editor    = {Ballester, P. and Egret, D. and Lorente, N. P. F.},
  url       = {https://arxiv.org/abs/1111.3618}
}

@article{Belter2015,
  author  = {Belter, Christopher W.},
  title   = {Bibliometric indicators: opportunities and limits},
  journal = {Journal of the Medical Library Association},
  volume  = {103},
  number  = {4},
  pages   = {219--221},
  year    = {2015},
  doi     = {10.3163/1536-5050.103.4.014}
}

@article{Waltman2016,
  author  = {Waltman, Ludo},
  title   = {A review of the literature on citation impact indicators},
  journal = {Journal of Informetrics},
  volume  = {10},
  number  = {2},
  pages   = {365--391},
  year    = {2016},
  doi     = {10.1016/j.joi.2016.02.007}
}

@article{Hicks2015,
  author  = {Hicks, Diana and Wouters, Paul and Waltman, Ludo and de Rijcke,
             Sarah and Rafols, Ismael},
  title   = {The Leiden Manifesto for research metrics},
  journal = {Nature},
  volume  = {520},
  pages   = {429--431},
  year    = {2015},
  doi     = {10.1038/520429a}
}

@article{Radicchi2008,
  author  = {Radicchi, Filippo and Fortunato, Santo and Castellano, Claudio},
  title   = {Universality of citation distributions: Toward an objective
             measure of scientific impact},
  journal = {Proceedings of the National Academy of Sciences},
  volume  = {105},
  number  = {45},
  pages   = {17268--17272},
  year    = {2008},
  doi     = {10.1073/pnas.0806977105}
}

@article{Thelwall2016,
  author  = {Thelwall, Mike},
  title   = {Citation count distributions for large monodisciplinary journals},
  journal = {Journal of Informetrics},
  volume  = {10},
  number  = {3},
  pages   = {863--874},
  year    = {2016},
  doi     = {10.1016/j.joi.2016.07.006}
}

@article{Casey2025effect,
	author = {Casey, Andy and Mandel, Ilya and Ray, P.K.},
	journal = {Bulletin of the AAS},
	number = {1},
	year = {2025},
	month = {aug 23},
	note = {https://baas.aas.org/pub/2025i020},
	publisher = {},
	title = {The effect of the {COVID}-19 pandemic on {arXiv} pre-prints in maths, physics, computer science, and quantitative biology},
	volume = {57},
}

@article{Lammey2020,
  author  = {Lammey, Rachael},
  title   = {Solutions for identification problems: a look at the Research
             Organization Registry},
  journal = {Science Editing},
  volume  = {7},
  number  = {1},
  pages   = {65--69},
  year    = {2020},
  doi     = {10.6087/kcse.192}
}

@misc{zenodo,
  author       = {{European Organization for Nuclear Research} and
                  {OpenAIRE}},
  title        = {Zenodo},
  year         = {2013},
  doi          = {10.25495/7GXK-RD71},
  url          = {https://zenodo.org/}
}

@article{
Szalay2001,
author = {Szalay and Gray 2001},
title = {The World-Wide Telescope},
journal = {Science},
volume = {293},
number = {5537},
pages = {2037-2040},
year = {2001},
doi = {10.1126/science.293.5537.2037},
URL = {https://www.science.org/doi/abs/10.1126/science.293.5537.2037},
eprint = {https://www.science.org/doi/pdf/10.1126/science.293.5537.2037},
}

@inproceedings{Hanisch2004,
  author    = {Quinn, Peter J. and others},
  title     = {The International Virtual Observatory Alliance:
               recent technical developments and the road ahead},
  booktitle = {Optimizing Scientific Return for Astronomy through
               Information Technologies},
  series    = {Proc.\ SPIE},
  volume    = {5493},
  pages     = {137--145},
  year      = {2004},
  doi       = {10.1117/12.551349}
}

@misc{accomazzi2014unifiedastronomythesaurus,
      title={The Unified Astronomy Thesaurus}, 
      author={Alberto Accomazzi and Norman Gray and Chris Erdmann and Chris Biemesderfer and Katie Frey and Justin Soles},
      year={2014},
      eprint={1403.6656},
      archivePrefix={arXiv},
      primaryClass={astro-ph.IM},
      url={https://arxiv.org/abs/1403.6656}, 
}

@article{Nelson2019,
  author  = {Nelson, Dylan and Springel, Volker and Pillepich, Annalisa and
             Rodriguez-Gomez, Vicente and Torrey, Paul and Genel, Shy and
             Vogelsberger, Mark and Pakmor, R{\"u}diger and Marinacci,
             Federico and Weinberger, Rainer and Kelley, Luke and
             Lovell, Mark and Diemer, Benedikt and Hernquist, Lars},
  title   = {The {IllustrisTNG} simulations: public data release},
  journal = {Computational Astrophysics and Cosmology},
  volume  = {6},
  pages   = {2},
  year    = {2019},
  doi     = {10.1186/s40668-019-0028-x}
}

@article{Evans2010,
  author  = {Evans, Ian N. and Primini, Francis A. and Glotfelty, Kenny J.
             and Anderson, Craig S. and Bonaventura, Nina R. and Chen,
             Judy C. and Davis, John E. and Doe, Stephen M. and Evans,
             Janet D. and Fabbiano, Giuseppina and others},
  title   = {The {Chandra} Source Catalog},
  journal = {The Astrophysical Journal Supplement Series},
  volume  = {189},
  number  = {1},
  pages   = {37--82},
  year    = {2010},
  doi     = {10.1088/0067-0049/189/1/37}
}

@article{Mann1947,
  author  = {Mann, Henry B. and Whitney, Donald R.},
  title   = {On a Test of Whether One of Two Random Variables is
             Stochastically Larger than the Other},
  journal = {The Annals of Mathematical Statistics},
  volume  = {18},
  number  = {1},
  pages   = {50--60},
  year    = {1947},
  doi     = {10.1214/aoms/1177730491}
}

@article{Kruskal1952,
  author  = {Kruskal, William H. and Wallis, W. Allen},
  title   = {Use of Ranks in One-Criterion Variance Analysis},
  journal = {Journal of the American Statistical Association},
  volume  = {47},
  number  = {260},
  pages   = {583--621},
  year    = {1952},
  doi     = {10.1080/01621459.1952.10483441}
}

@article{MacKinnon1985,
  author  = {MacKinnon, James G. and White, Halbert},
  title   = {Some heteroskedasticity-consistent covariance matrix estimators
             with improved finite sample properties},
  journal = {Journal of Econometrics},
  volume  = {29},
  number  = {3},
  pages   = {305--325},
  year    = {1985},
  doi     = {10.1016/0304-4076(85)90158-7}
}

@book{Kendall1948,
  author    = {Kendall, Maurice G.},
  title     = {Rank Correlation Methods},
  publisher = {Griffin},
  address   = {London},
  year      = {1948}
}

@article{Schaye2015,
  author  = {Schaye, Joop and Crain, Robert A. and Bower, Richard G. and
             Furlong, Michelle and Schaller, Matthieu and Theuns, Tom and
             Dalla Vecchia, Claudio and Frenk, Carlos S. and McCarthy,
             Ian G. and Helly, John C. and Jenkins, Adrian and
             Rosas-Guevara, Y.~M. and White, Simon D.~M. and Baes,
             Maarten and Booth, C.~M. and Camps, Peter and Navarro,
             Julio F. and Qu, Yan and Rahmati, Alireza and Sawala, Till
             and Thomas, Peter A. and Trayford, James},
  title   = {The {EAGLE} simulations of galaxy formation: calibration of
             subgrid physics and model variations},
  journal = {Monthly Notices of the Royal Astronomical Society},
  volume  = {446},
  pages   = {521--554},
  year    = {2015},
  doi     = {10.1093/mnras/stu2058}
}

@misc{python3,
  author       = {{Python Core Team}},
  title        = {Python: A dynamic, open source programming language},
  year         = {2023},
  howpublished = {\url{https://www.python.org/}},
  note         = {Python Software Foundation}
}

@article{numpy,
  author  = {Harris, Charles R. and Millman, K. Jarrod and van der Walt,
             Stefan J. and Gommers, Ralf and Virtanen, Pauli and
             Cournapeau, David and Wieser, Eric and Taylor, Julian and
             Berg, Sebastian and Smith, Nathaniel J. and others},
  title   = {Array programming with {NumPy}},
  journal = {Nature},
  volume  = {585},
  pages   = {357--362},
  year    = {2020},
  doi     = {10.1038/s41586-020-2649-2}
}

@inproceedings{pandas,
  author    = {McKinney, Wes},
  title     = {Data Structures for Statistical Computing in {Python}},
  booktitle = {Proceedings of the 9th Python in Science Conference},
  pages     = {56--61},
  year      = {2010},
  doi       = {10.25080/Majora-92bf1922-00a}
}

@article{matplotlib,
  author  = {Hunter, John D.},
  title   = {Matplotlib: A 2{D} Graphics Environment},
  journal = {Computing in Science \& Engineering},
  volume  = {9},
  number  = {3},
  pages   = {90--95},
  year    = {2007},
  doi     = {10.1109/MCSE.2007.55}
}

@article{scipy,
  author  = {Virtanen, Pauli and Gommers, Ralf and Oliphant,
             Travis E. and Haberland, Matt and Reddy, Tyler and
             Cournapeau, David and Burovski, Evgeni and Peterson, Pearu
             and Weckesser, Warren and Bright, Jonathan and others},
  title   = {{SciPy} 1.0: Fundamental Algorithms for Scientific
             Computing in {Python}},
  journal = {Nature Methods},
  volume  = {17},
  pages   = {261--272},
  year    = {2020},
  doi     = {10.1038/s41592-019-0686-2}
}

@inproceedings{statsmodels,
  author    = {Seabold, Skipper and Perktold, Josef},
  title     = {Statsmodels: Econometric and statistical modeling
               with {Python}},
  booktitle = {Proceedings of the 9th Python in Science Conference},
  pages     = {92--96},
  year      = {2010},
  doi       = {10.25080/Majora-92bf1922-011}
}

@misc{Momcheva2015,
      title={Software Use in Astronomy: an Informal Survey}, 
      author={Ivelina Momcheva and Erik Tollerud},
      year={2015},
      eprint={1507.03989},
      archivePrefix={arXiv},
      primaryClass={astro-ph.IM},
      url={https://arxiv.org/abs/1507.03989}, 
}

@misc{mycode, 
    author = {Joshi and Croft},
    title = {Code Repository for Open Science in Astrophysics: Citation Benefits of Open Code, Open Data, and Open Access},
    year = {2026},
    url = {https://github.com/hellothere98765/Code-For-Accessibility}}

@misc{mydata,
    author = {Joshi and Croft},
    title = {Dataset for Open Science in Astrophysics: Citation Benefits of Open Code, Open Data, and Open Access},
    year = {2026},
    publisher = {Zenodo},
    doi = {10.5281/zenodo.20270191}
    }

\end{document}